\documentclass[12pt,english,aps,tightenlinesletterpaper,groupaddress,secnumarabic,nofootinbib]{revtex4}
\usepackage{amsmath}
\usepackage{mathptmx}
\usepackage{newtxmath}

\usepackage[T1]{fontenc}
\usepackage[latin9]{inputenc}
\usepackage[a4paper]{geometry}
\usepackage{babel}
\usepackage{float}
\usepackage{graphicx}
\usepackage[unicode=true,pdfusetitle,
 bookmarks=true,bookmarksnumbered=false,bookmarksopen=false,
 breaklinks=false,pdfborder={0 0 1},backref=false,colorlinks=false]
 {hyperref}

\makeatletter

\providecommand{\tabularnewline}{\\}
\@ifundefined{textcolor}{}
{%
 \definecolor{BLACK}{gray}{0}
 \definecolor{WHITE}{gray}{1}
 \definecolor{RED}{rgb}{1,0,0}
 \definecolor{GREEN}{rgb}{0,1,0}
 \definecolor{BLUE}{rgb}{0,0,1}
 \definecolor{CYAN}{cmyk}{1,0,0,0}
 \definecolor{MAGENTA}{cmyk}{0,1,0,0}
 \definecolor{YELLOW}{cmyk}{0,0,1,0}
}

\newcommand{\eft}{EFTofLSS }
\newcommand{\ctp}{counterterm parameters }

\usepackage{xcolor}

\makeatother

\begin{document}
%\title{Forecasting blue \moritz{keep?} isocurvature detectability in Euclid and MegaMapper using EFTofLSS}

\title{Search for Isocurvature with Large-scale Structure: A Forecast for Euclid and MegaMapper using EFTofLSS}

\author{Daniel J. H. Chung}
\email{danielchung@wisc.edu}

\affiliation{Department of Physics, University of Wisconsin-Madison, Madison, WI
53706, USA}
\author{Moritz M{\"u}nchmeyer}
\email{muenchmeyer@wisc.edu}

\affiliation{Department of Physics, University of Wisconsin-Madison, Madison, WI
53706, USA}
\author{Sai Chaitanya Tadepalli}
\email{stadepalli@wisc.edu}

\affiliation{Department of Physics, University of Wisconsin-Madison, Madison, WI
53706, USA}
\begin{abstract}
Isocurvature perturbations with a blue power spectrum are one of the natural targets for
the future large scale structure observations which are probing shorter
length scales with greater accuracy. We present a Fisher forecast
for the Euclid and MegaMapper (MM) experiments in their ability to detect
blue isocurvature perturbations. We construct the theoretical predictions in the EFTofLSS and bias expansion
formalisms at quartic order in overdensities which allows us to compute
the power spectrum at one loop order and bispectrum at tree level
and further include theoretical error at the next to leading order
for the covariance determination. We find that Euclid is expected
to provide at least a factor of few improvement on the isocurvature
spectral amplitude compared to the existing Planck constraints for
large spectral indices while MM is expected to provide about 1 to
1.5 order of magnitude improvement for a broad range of spectral indices.
We find features that are specific to the blue isocurvature scenario
including the leading parametric degeneracy being with the Laplacian
bias and a UV sensitive bare sound speed parameter.

\tableofcontents{}
\end{abstract}
\maketitle

\section{\label{sec:Introduction}Introduction}

Cold dark matter (CDM) isocurvature perturbations with a strongly
blue spectral tilt can be completely negligible on large scales and
be the dominant primordial inhomogeneity component on small length
scales. Cosmic microwave background (CMB) strongly constrains the
large scales and is currently the dominant anchor of standard cosmology
\citep{Planck:2015fie,Planck:2018jri}, but future large scale structure
surveys are poised to exceed the constraining power of the CMB on
short length scales \citep{Sailer:2021yzm}. In this paper, we give
a Fisher forecast of the constraining power of Euclid \citep{Euclid}
and MegaMapper (MM) \citep{Schlegel:2019eqc} on strongly blue tilted
isocurvature spectra.

CDM blue isocurvature perturbations are naturally generated in axion-like
scenarios when the Peccei-Quinn symmetry breaking radial field is
out of equilibrium during inflation \citep{Kasuya:2009up}. The duration
of the time that the radial field is not at the minimum of the effective
potential determines the $k$ range over which the spectrum is blue,
with the spectral index being determined by the logarithmic time derivative
of the radial field. In the overdamped radial dynamics, the spectrum
is typically characterized by an approximately constant spectral index
until the break point when the spectrum becomes flat again. While
the overdamped scenario has only a mild bump at the break point \citep{Chung:2016wvv},
the underdamped scenarios can generate spectacular features near the
break point that can include oscillations of huge amplitudes \citep{Chung:2021lfg}.
In the present work, we consider a simplified version of these more
physically complete scenarios by restricting to a single power law
characterized by an amplitude and a spectral index.

For the theoretical cosmological fluid model that will be used to
compute experimental observables, we use the \eft formalism \citep{Baumann:2010tm}
at one-loop order because it provides a principled way to control uncertainties from unknown small-scale physics in the spectral range of interest. We calibrate the counterterm of \eft using a combination of codes CLASS-PT \citep{CLASS_PT},
FastPM \citep{Feng:2016yqz}, and nbodykit \citep{Hand:2017pqn}.
We model the galaxy counts using a bias expansion \citep{Desjacques_2018}
and compute the galaxy power spectrum at one-loop order and the bispectrum
at tree-level. We include theoretical error estimates for the covariance
determination using the higher-order error envelopes as given in \citep{Baldauf:2016sjb,Chudaykin:2019ock}.

We find that Euclid can give a factor of a few improvement on the isocurvature
spectral amplitude constraint over that from the CMB for large isocurvature
spectral indices ($n_{\mathrm{iso}}\gtrsim3$), while MM can give more
than an order of magnitude improvement for a broad range of $n_{\mathrm{iso}}$.
The controlling factor in the degree of constraint is the signal data
volume, and most of the signal (at the perturbative order used in
this work) is coming from the power spectrum and not the bispectrum
for these experiments, although bispectrum signal helps to break some
parameter degeneracies. We identify the dominant source of signal
degradation to come from the marginalization associated with the Laplacian
bias due to a degeneracy of high spectral indexed blue isocurvature
spectra with this Laplacian bias $k$-dependence.

One feature of the strongly blue isocurvature scenario in the \eft
formalism that we uncover is that the bare sound speed parameter $c^{2}$
is both cutoff dependent and may even become negative when the cutoff
is taken above $\Lambda\gtrsim O(3)$ Mpc$^{-1}$ for sizable isocurvature
amplitudes. This does not significantly affect the analysis because
the renormalized $c_{\mathrm{ren}}^{2}$ is of the same approximate
value as in the case of the adiabatic scenarios. Along a similar theme,
the one loop contributions to the galaxy power spectrum have integrals
that are UV-sensitive ($\Lambda$-dependent) for high spectral indices.
Again, we have enough bias parameter renormalization degree of freedom
to absorb the UV sensitivities and marginalize, such that these peculiarities
do not strongly affect the Fisher forecast in practice.

The order of presentation is as follows. In Sec.~\ref{sec:Mixed-power-spectrum},
we parameterize the mixed (adiabatic + isocurvature) linear input power spectra. We review the
\eft formalism in Sec.~\ref{sec:Review-of-eft} and discuss the
one-loop \eft parameterization used for the forecast. Here we also
discuss the enhanced UV-sensitivity of the loop integrals coming from
the large spectral index of the linear isocurvature power spectrum.
In Sec.~\ref{sec:Galaxy-power-spectrum}, we present the bias expansion
used to model the galaxy power spectrum and bispectrum. Sec.~\ref{sec:UV-div-and-corr}
discusses the divergences and the renormalization procedure in the
bias expansion. In Sec.~\ref{sec:Parameter-set-and}, we give the
assumed characteristics associated with the Euclid and MM experiments
as well as the fiducial parameter set associated with the cosmology.
We discuss the results in Sec.~\ref{sec:Results-and-Discussion}
where the Euclid and MM sensitivity to strongly blue isocurvature
spectra is presented. We conclude the paper in Sec.~\ref{sec:Conclusion}
with a summary of main results and an outlook of future related work.

The set of appendices that follow present some of the technical details.
In App.~\ref{sec:Renormalization-scheme}, the renormalization scheme used
in this paper is defined. We list the one loop integrals appearing
in the bias expansion in App.~\ref{sec:Galaxy-one-loop-contributions}.
App.~\ref{sec:Semi-analytic-expression-for-csrenMX} gives the renormalized
$c^{2}$ for the mixed case as a function of $z$ obtained by matching with N-body simulations.
For a self-contained presentation, we review the Fisher forecast formalism
and list our simplifying assumptions in App.~\ref{sec:Fisher-matrix-and}.  In App.~\ref{sec:halofit} we discuss how the current version of Halofit \cite{Takahashi:2012em} is insufficient to describe the nonlinear power spectrum for the mixed initial condition scenarios.

\section{\label{sec:Mixed-power-spectrum}Mixed power spectrum scenario}

We work in the standard 6-parameter $\Lambda$CDM cosmology with the
addition of axion dark matter component with isocurvature initial
conditions (ICs). Axion-like scenarios can naturally generate a large
spectral tilt (strongly blue) CDM isocurvature perturbations when
the Peccei-Quinn symmetry breaking radial field is out of equilibrium
during inflation \citep{Kasuya:2009up}. The existence of a break
in the spectrum is generic and required for the observability of the
spectrum \citep{Chung:2015tha}. Furthermore, different initial conditions
can lead to a rich set of dynamics that is seen in the features of
the spectrum \citep{Chung:2016wvv,Chung:2021lfg}. In the case where
the break point of the spectrum is far beyond $k_{\mathrm{NL}}$,
we expect the perturbative observables to be sensitive mainly to the power law
part of the blue spectrum. Hence, deferring a more complete analysis
to a future work, we consider a simplified scenario in this paper
and parameterize the primordial curvature and isocurvature spectra
as the following standard power-law expressions (by introducing the
dimensionless power spectrum $\Delta_{X}^{2}(k)=k^{3}P_{X}(k)/\left(2\pi^{2}\right)$)
\begin{equation}
\frac{k^{3}}{2\pi^{2}}P_{\mathcal{R}}(k)=A_{{\rm ad}}(k_{\mathrm{p}})\left(\frac{k}{k_{\mathrm{p}}}\right)^{n_{\mathrm{ad}}-1},\qquad\frac{k^{3}}{2\pi^{2}}P_{\mathcal{S}_{cdm}}(k)=A_{{\rm iso}}(k_{\mathrm{p}})\left(\frac{k}{k_{\mathrm{p}}}\right)^{n_{\mathrm{iso}}-1}
\end{equation}
at a pivot scale $k_{{\rm p}}=0.05\,{\rm Mpc^{-1}}$. Since axionic
theories give $1\leq n_{{\rm iso}}\leq4$ in the scaling part of the
spectrum, we restrict our $n_{{\rm iso}}$ to this range. We also
assume that the curvature-isocurvature cross-correlation is negligible
which corresponds to a situation where the axion coupling to the inflaton
is sufficiently suppressed. Due to the linearity of first order perturbation
theory, the general mixed perturbations evolve as a linear superposition
of the adiabatic (AD) and isocurvature (ISO) components that evolve
independently and hence the total linear matter power spectrum in
presence of both adiabatic and CDM-isocurvature fluctuations (mixed
(MX) state) can be given as
\begin{align}
P_{{\rm lin}}^{{\rm MX}}(k,z\ll z_{{\rm eq}}) & =P_{{\rm lin}}^{{\rm {\rm AD}}}(k,z)+P_{{\rm lin}}^{{\rm ISO}}(k,z).
\end{align}
This then allows us to write the linear power spectrum at the time
when the perturbative loop expansion will be made as
\begin{align}
P_{{\rm lin}}^{{\rm MX}}(k,z\ll z_{{\rm eq}}) & =P_{{\rm lin}}^{{\rm {\rm AD}}}(k,z)\left(1+\alpha\left(\frac{f_{c}}{3}\right)^{2}\left(\frac{T_{{\rm iso}}(k)}{T_{{\rm ad}}(k)}\right)^{2}\left(\frac{k}{k_{\mathrm{p}}}\right)^{n_{{\rm iso}}-n_{{\rm ad}}}\right)\label{eq:Pm_analytical}
\end{align}
where \textbf{
\begin{equation}
\alpha=\frac{A_{{\rm iso}}(k_{{\rm p}})}{A_{{\rm ad}}(k_{{\rm p}})}
\end{equation}
}and $f_{c}$ is the fraction of cold dark matter (CDM) energy density
over total dark matter and baryon energy density assuming that the
primordial isocurvature mode is generated within CDM fluctuations.
The additional factor of $\left(1/3\right)^{2}$ is obtained by mapping
superhorizon isocurvature modes to curvature perturbations during
Matter-Dominated (MD) era due to the change in the effective sound
speed-squared\footnote{
\begin{equation}
\mathcal{R}_{k}=\mathcal{R}_{k}({\rm rad)}+\frac{y}{4+3y}\mathcal{S}_{k}({\rm rad)}\qquad\mbox{for \ensuremath{k\ll\mathcal{H}}}\label{eq:isocurv_to_curv}
\end{equation}
where $y=a/a_{{\rm eq}}$ such that the effective sound speed-squared
for combined matter-radiation fluid, $c_{s}^{2}=4/3/\left(4+3y\right)$,
and thus the prefactor $y/\left(4+3y\right)\rightarrow1/3$ for $y\rightarrow\infty$
in Eq.~(\ref{eq:isocurv_to_curv}) \citep{osti_5802428,Hannu}.} $c_{s}^{2}$. The transfer functions $T_{{\rm iso}}(k)$ and $T_{{\rm ad}}(k)$
are standard linear transfer functions associated with the $\Lambda$CDM
cosmology. By construction, $T\rightarrow1$ as $k\rightarrow0$ (for
very large scales) whereas it decreases monotonically at small scales.
The actual shape dependence of $T(k)$ relies upon the primordial
mode (adiabatic or isocurvature) sourcing the fluctuations where isocurvature
modes suffer from additional small scale suppression compared to adiabatic
since they are characterized as initial zero energy density or potential
modes \citep{1986ApJ,Efstathiou:1986pba}. 
\begin{figure}
\begin{centering}
\includegraphics[scale=0.6]{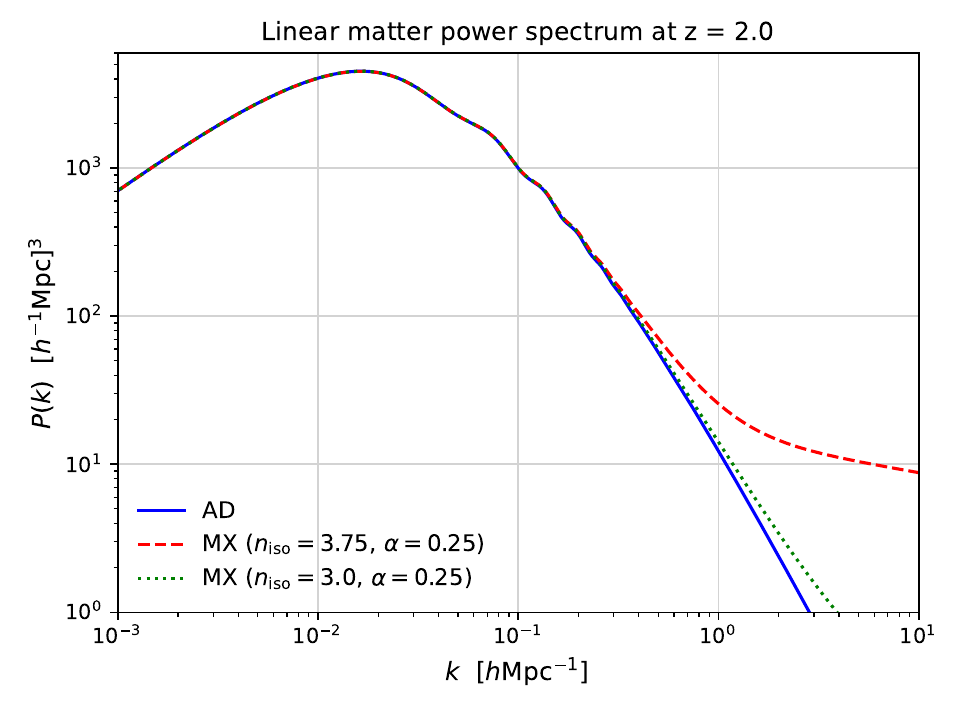}
\par\end{centering}
\caption{\label{fig:Plin}Plot showing a comparison between the linear matter power spectra at a redshift of $z=2$ for the pure adiabatic and mixed initial conditions. For the mixed case, we use fiducial values as $\alpha=0.25$ and plot two curves for  $n_{\mathrm{iso}}=3.75$ (dashed) and  $n_{\mathrm{iso}}=3.0$ (dotted).}
\end{figure}
To illustrate the comparison between the two initial conditions, in Fig.~\ref{fig:Plin} we plot linear matter power spectrum at a redshift of $z=2$ generated from the CLASS code \citep{Blas:2011rf} for the pure adiabatic and mixed initial conditions. To gain a perspective into
the magnitude of $\alpha$, we note that latest Planck analysis constraints
$\alpha\lesssim O(0.1)$ \citep{Planck:2018jri} at $k_{{\rm p}}$
which implies that $P_{{\rm lin}}^{{\rm ISO}}(k_{{\rm p}})\lesssim O(0.01)P_{{\rm lin}}^{{\rm AD}}(k_{{\rm p}})$.
Hence, isocurvature power is roughly $O\left(1\%\right)$ or less
than adiabatic on large scales.

\section{\label{sec:Review-of-eft}The \eft with isocurvature}
%\moritz{changed title}

In this work we use the \eft to model the galaxy power spectrum and bispectrum observables on perturbative non-linear scales. Although cosmic emulators such as \citep{Moran:2022iwe},
seem to fit the numerical simulations well for the adiabatic case over a wide range for the cosmological parameters, the current versions are not trained for mixed isocurvature and are also subject to baryonic uncertainty. The semi-analytical Halofit approach \citep{Takahashi:2012em} to clustering also fails when one includes isocurvature perturbations, as we briefly explore in App. \ref{sec:halofit}. On the other hand, the \eft \citep{Baumann:2010tm,Carrasco_2012,Ivanov:2022mrd}
has a well-defined expansion parameter that controls errors and can be extended to include isocurvature. The EFT approach
models UV-effects/backreaction of small-scale non-linear physics through
various unknown free parameters whose values are determined from simulations.

\eft in the mixed case has a mild disadvantage compared to the adiabatic
case because of the relatively larger variation in the counterterm
fitting parameter such as renormalized $c^{2}$ with the changes in
the isocurvature model parameters. This is because the blue isocurvature spectrum becomes large on small scales, which critically affects the UV behavior of the theory. However, because of the degeneracy
of these parameters with the bias parameters and because the theory
error is dominated by the next order term in the loop expansion, this
variation in the \ctp does not play a significant role in the current
analysis. 

In the present forecast, we will consider our theoretical model for
power spectrum and bispectrum by including all contributions up to
quartic power in linear matter overdensity, $O\left(\left(\delta^{(1)}\right)^{4}\right)$,
where throughout our analysis we will assume Gaussian initial conditions
such that $\left\langle \delta^{(1)}\right\rangle =0$. At this order
in perturbative expansion, we can at most work with one-loop galaxy
power spectrum and tree-level bispectrum \citep{Angulo:2015eqa}.
As we will see, this is sufficient to see a signal in the experiments
that we forecast.

As done in \citep{Carrasco_2012}, one can expand the non-linear matter
overdensity perturbatively to one-loop as
\begin{equation}
\delta=\delta^{(1)}+\delta^{(2)}+\delta^{(3)}+\delta^{(c_{1})}+\epsilon_{{\rm m}}\label{eq:matterexpansion}
\end{equation}
where $\delta^{(1,2,3)}$ are effectively Standard Perturbation Theory
(SPT \citep{Bernardeau:2001qr}) solutions (i.e. $\delta_{{\rm SPT}}\approx\delta^{(1)}+\delta^{(2)}+\delta^{(3)}$
to third order in $\delta^{(1)}$), and $\delta^{(c_{1})}$ and $\epsilon_{{\rm m}}$
are the one-loop counterterm and stochastic noise term.\footnote{By $\delta^{(n)}$ being effective SPT solutions, we mean that these
formally have window functions attached to them coming from the smearing
operation, but on scales of interest, these evaluate to unity such
that the smeared solutions are the same as the SPT solutions.} More specifically, the terms $\delta^{(c_{1})}+\epsilon_{{\rm m}}$
represent the corrections in the \eft arising from residual pressure
sources associated with averaging the Euler equation, where these
sources have been expanded in powers of derivatives and $\delta^{(1)}$
with the expansion coefficients ( such as $c^{2}$) parameterized
instead of being computed. This parameterization allows an introduction
of a $\Lambda$-dependent coefficient that can be used to remove the
$\Lambda$-dependence in the observable correlator. The often called
stochastic term $\epsilon_{{\rm m}}$ is defined to be independent
of $\delta^{(1)}$ such that its contribution is uncorrelated with
$\delta^{(n)}$, e.g.~$\left\langle \delta^{(1)}\epsilon_{{\rm m}}\right\rangle =0$.
Using this expansion, the following standard expression for the non-linear
power spectrum is derived in \citep{Carrasco_2012} as
\begin{equation}
P_{{\rm NL}}(k,z)=P_{{\rm 11}}(k,z)+\left[P_{{\rm 22}}(k,z)+P_{13}(k,z)\right]+P_{{\rm ctr}}(k,z)+P_{\Delta\tau}(k,z),\label{eq:1-loopPmmodel}
\end{equation}
where $P_{{\rm 11}}\equiv P_{{\rm lin}}$ is the linear matter power
spectrum, and that $P_{{\rm 12}}(k,z)$ vanishes due to our assumption
of Gaussian initial conditions. In the \eft, although the quantity
$P_{11}$ is the power spectrum of the fields smeared over length
scales shorter than $\Lambda^{-1}$, because $P_{11}$ is evaluated
at $k\ll\Lambda$, we approximate $P_{11}$ as the unsmeared power
spectrum.

The one-loop contributions $P_{22}$ and $P_{13}$ are given by the
expressions
\begin{equation}
P_{22}(k)=2\int_{q}F_{2}^{2}\left(\bar{q},\bar{k}-\bar{q}\right)P_{{\rm 11}}(q)P_{{\rm 11}}(|\bar{k}-\bar{q}|),\label{eq:P22_0}
\end{equation}
\begin{equation}
P_{13}(k)=6P_{{\rm 11}}(k)\int_{q}F_{3}\left(\bar{q},-\bar{q},\bar{k}\right)P_{{\rm 11}}(q),\label{eq:P13_0}
\end{equation}
where $\int_{q}\equiv\int\frac{d^{3}q}{\left(2\pi\right)^{3}}$ and
$\bar{k}=\bar{q}_{1}+...+\bar{q}_{n}$. The explicit expressions for
the standard SPT convolution kernels $F_{2,3}$ are given in \citep{Bernardeau_2002}
whereas simplified loop integrals useful for all evaluations can be
found in \citep{Carrasco_2012,Senatore:2014via}. The SPT expressions for
$P_{22}$ and $P_{13}$ in Eqs.~(\ref{eq:P22_0}) and (\ref{eq:P13_0})
are correct only in an Einstein de-Sitter (EdS) universe where momentum
and time-dependence of the convolution loop integrals decouple. However,
it is a common approach to use EdS based kernels $F_{2,3}$ even in
non-EdS spacetimes and subsequently replace the linear growth function,
$D_{+}(z)$, consistent with non-EdS cosmologies because the momentum
addition deformations are not as important as the scaling from the
growth function. This approach has been tested analytically \cite{Takahashi:2008yk} and with a number of N-body
simulations \citep{Baldauf_2015} and found to be very accurate such that the residual difference
is negligible for foreseeable future surveys.

\begin{figure}
\begin{centering}
\includegraphics[scale=0.5]{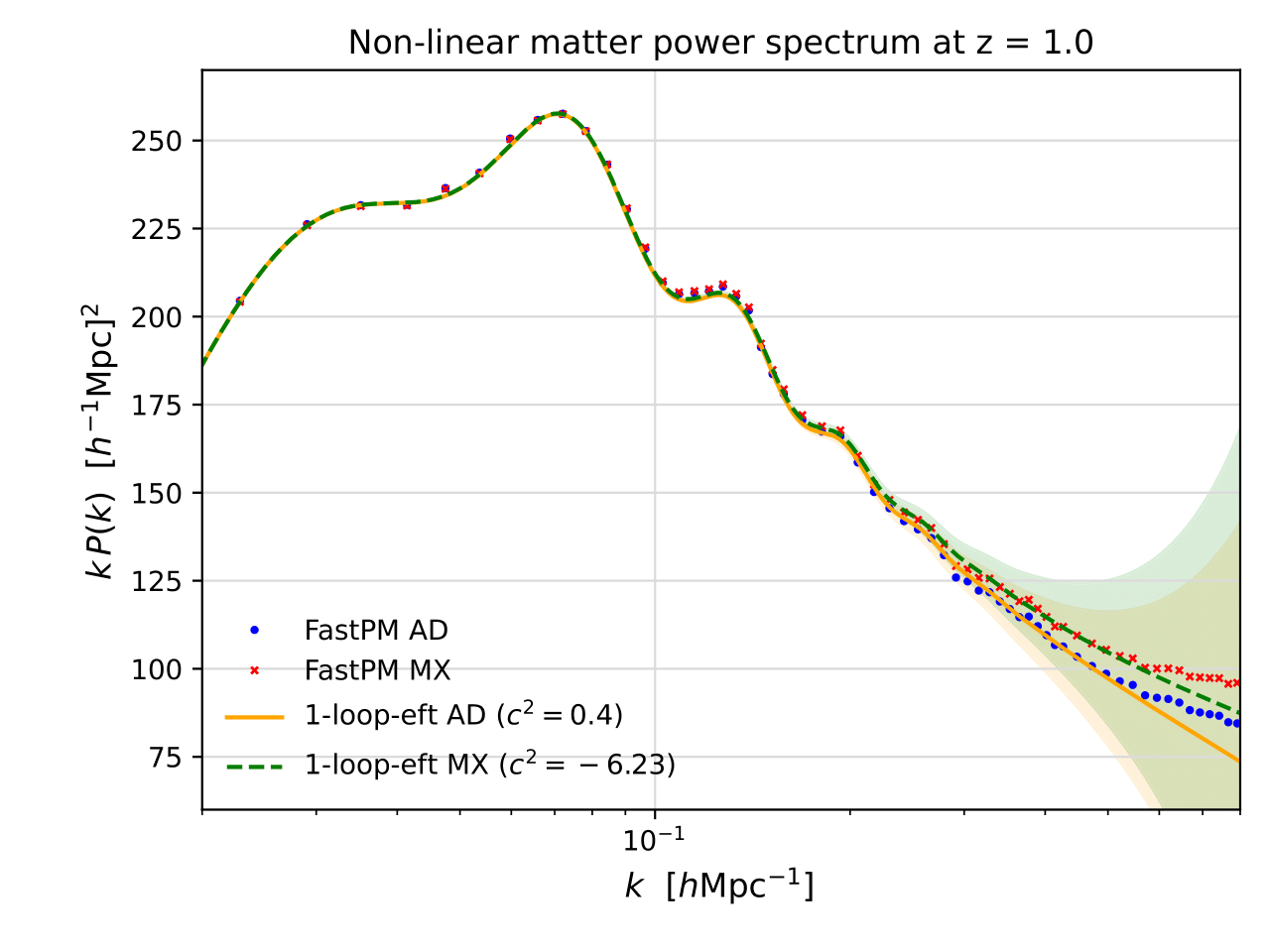}~~\includegraphics[scale=0.5]{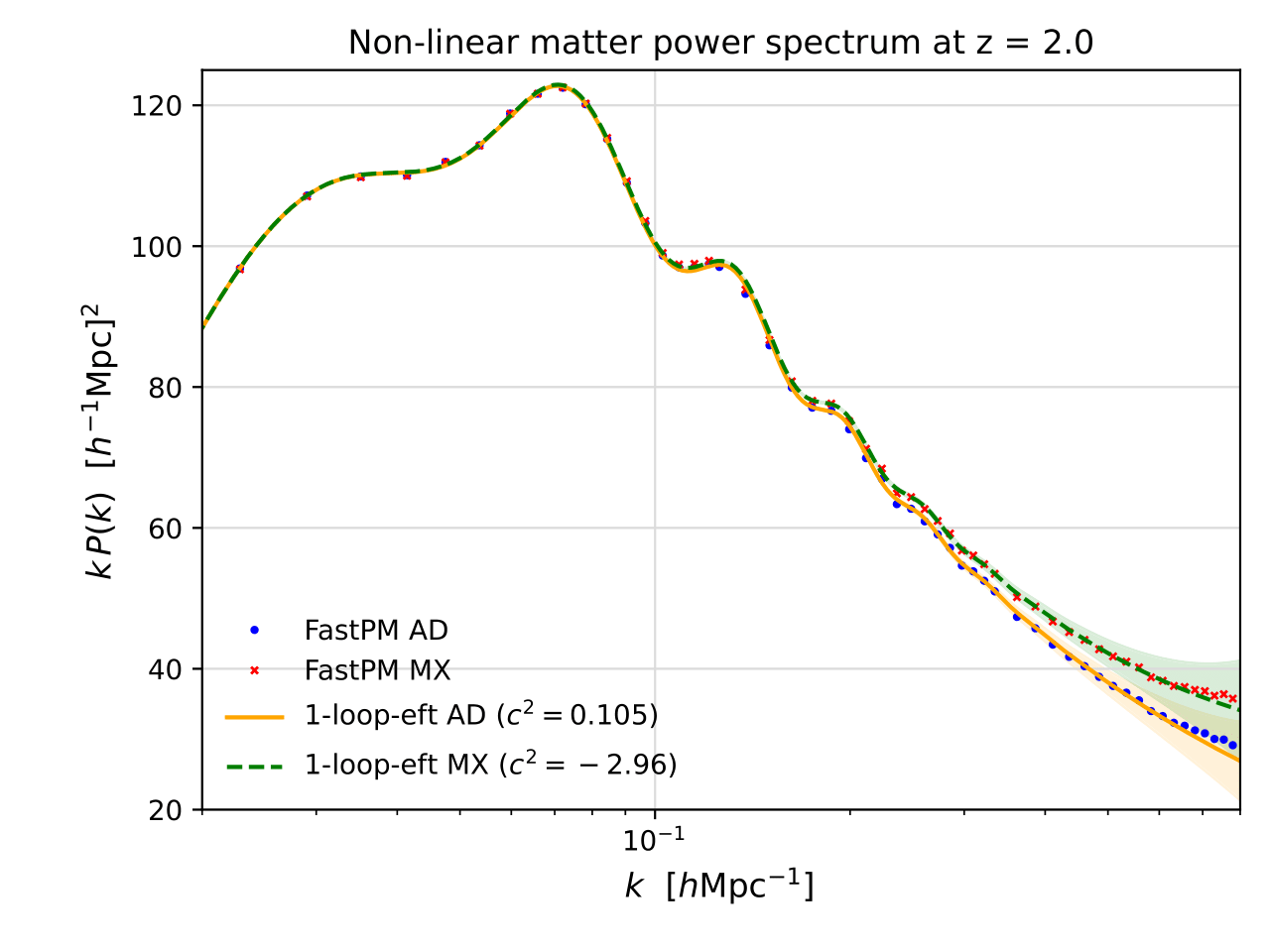}
\par\end{centering}
\caption{\label{fig:eft-fastpm-compare}In this figure we highlight the fitting of one-loop EFT power spectrum to the N-body data from FastPM \cite{Feng:2016yqz}. Note that we plot scaled power spectrum, $k\times P(k)$, on the y-axis for clarity. For the mixed case, we use fiducial values as $n_{\rm{iso}}=3.75$ and $\alpha=0.25$. The value of the bare $c^2$ (at cutoff $\Lambda=100$ h/Mpc) one-loop EFT parameter is given in the label for the EFT curves. Note that the value of $c^2$ for the mixed case is negative. We discuss this peculiarity in Sec.~\ref{subsec:peculiar}. The $k_{\mathrm{reach}}$ of the one-loop EFT curves is $\approx$ $0.5~$h/Mpc ($0.65~$h/Mpc) at redshift $z=1$ ($z=2$). We also plot approximate theoretical error band expected from two-loop contributions (see Eq.~(\ref{eq:error}). As can be seen in the plots, the two-loop error estimate is quite conservative. See Fig.~\ref{fig:Plin} to compare with the corresponding linear power spectra. In Appendix \ref{sec:halofit} we discuss comparison of N-body results with Halofit and CosmicEmu.}
\end{figure}

The counterterm at the one-loop order (coming from $\left\langle \delta^{(c_{1})}\delta^{(1)}\right\rangle $)
is conveniently represented as 
\begin{equation}
P_{{\rm ctr}}(k,z)=-2c^{2}(z)k^{2}P_{11}(k,z)\label{eq:Pctr}
\end{equation}
where $c^{2}$ is an effective parameter arising from the parameterization
of $\delta^{(c_{1})}$ \footnote{$\delta^{(c_{1})}=c^{2}\nabla^{2}\delta^{(1)}$ \citep{Carrasco_2012}}
. The higher order $k$ terms not shown in Eq.~(\ref{eq:Pctr}) will
represent higher than 4th perturbative order since we can view $k^{2}$
as second order in $k/k_{\mathrm{NL}}$ and $P_{{\rm 11}}(k,z)\propto\left(\delta^{(1)}\right)^{2}$
as second perturbative order. The parameter $c^{2}$ models the effective
sound speed-squared that arises due to the gravitational clustering
induced by small-scale fluctuations \citep{Carrasco_2012,Baldauf_2015}. It is noteworthy that the spectral shape of the contribution in Eq.~\ref{eq:Pctr} also accounts for the leading order effects from baryonic physics on large-scales \cite{Lewandowski:2014rca}, because total matter over-density $\delta$ is sum of cdm and baryonic contributions. These effects are most important post star-formation and their contribution is modeled effectively through the modification of the parameter $c^{2}$. As we shall note in Sec.~\ref{sec:Galaxy-power-spectrum}, the degeneracy of $c^{2}$ with the Laplacian bias parameter $b_{\nabla^2 \delta}$ implies we can effectively account for these by treating Laplacian bias as a free nuisance parameter.
In practice, $c^{2}$ is determined by matching Eq.~(\ref{eq:1-loopPmmodel})
to numerical simulations. If $P_{13}$ is divergent with a cutoff
dependent term having the same $k$ functional form as Eq.~(\ref{eq:Pctr}),
then $c^{2}$ is necessarily divergent (i.e. cutoff dependent) and
its non-divergent piece after combining with $P_{13}$ is the observationally
relevant ``renormalized'' piece, which we will call $c_{{\rm ren}}^{2}$ \footnote{One can factorize the bare parameter $c^2(\Lambda,z)$ into a renormalized part, $c_{\rm{ren}}^{2}(k_{\rm{ren}},z)$, and a counterterm part, $c_{{\rm \Lambda}}^{2}(k_{\rm{ren}},\Lambda,z)$ as  
\begin{equation}
c^{2}(\Lambda,z)=c_{{\rm \Lambda}}^{2}(k_{\rm{ren}},\Lambda,z)+c_{\rm{ren}}^{2}(k_{\rm{ren}},z)
\end{equation}
where $k_{\rm{ren}}$ is a scale chosen to set the value of the counterterm $c_{{\rm \Lambda}}^{2}(k_{\rm{ren}},\Lambda,z)$ \cite{Hertzberg_2014}. See Appendix \ref{sec:Renormalization-scheme} for details.
}
:
i.e. we can rewrite Eq.~(\ref{eq:1-loopPmmodel}) as
\begin{align}
P_{{\rm NL}}(k,z) &= P_{{\rm 11}}(k,z)+P_{{\rm 22}}(k,z)+\left(P_{13}(k,\Lambda,z)-2c_{{\rm \Lambda}}^{2}(k_{\rm{ren}},\Lambda,z)k^{2}P_{{\rm 11}}(k,z)\right) \nonumber \\ & -2c_{{\rm ren}}^{2}(k_{\rm{ren}},z)k^{2}P_{{\rm 11}}(k,z)  + P_{\Delta\tau}
(k,z)
\label{eq:renormalized_1loopPmm}
\end{align}
where $P_{11}$ and $P_{22}$ have a weak cutoff dependence for $k\ll\Lambda$, and $c_{{\rm \Lambda}}^{2}(k_{\rm{ren}},\Lambda,z)$ will have the cutoff dependence
to cancel the divergent (UV-sensitive) part of the $P_{13}$ term.
From dimensional analysis, one would expect the parameter $c_{\mathrm{ren}}^{2}(k_{\rm{ren}},z)$
to be of order $\sim1/k_{{\rm NL}}^{2}$. In the present work, we
use CLASS code \citep{Blas:2011rf} to generate the linear power spectra
which is input to FastPM \citep{Feng:2016yqz} whose N-body output
phase space data is processed by nbodykit \citep{Hand:2017pqn} to
generate the nonlinear power spectra to compute $c_{{\rm ren}}^{2}$. In Fig.~\ref{fig:eft-fastpm-compare} we show examples of the one-loop EFT matter power spectrum fitted to the N-body simulations obtained from FastPM\footnote{In \citep{Feng:2016yqz} the authors report that FastPM can achieve a $1\%$ level accuracy compared to TreePM solvers at $1$h/Mpc for at least 40 time-steps and a force resolution of about $2-3$. We chose $z_i = 99$ as the initial redshift for setting up the initial conditions and ran 1000 particles per box side for three different simulation boxes of side length $L=2$, $1$ and $0.512$ Gpc combining power spectrum data from each box such that the overall shot noise $<1\%$.} for both adiabatic and mixed ICs. The curves are plotted for redshifts $z=1$ and $z=2$. The $k_{\rm{reach}}$ which we define as the smallest mode where the EFT curve deviates by more than $1.5\%$ from N-body data is $\approx$ $0.5~$h/Mpc ($0.65~$h/Mpc) at redshift $z=1$ ($z=2$) respectively.

The additional contribution $P_{\Delta\tau}$ is obtained by the two-point
function of $\epsilon_{m}\sim k^{2}\Delta\tau$ where $\Delta\tau$
has an intuitive interpretation of residual effective pressure coming
from the smearing operation. Unlike $P_{{\rm ctr}}$, the stochastic
term $P_{\Delta\tau}$ scales as $k^{4}$ but is found to be negligible
for the purposes of our analysis because of the following argument.
We know $\Delta\tau$ has the magnitude of pressure, which means that
the two-point function of $\Delta\tau$ in $k$ space has a natural
magnitude of pressure squared divided by $k_{{\rm NL}}^{3}$. The
appearance of $k_{{\rm NL}}^{3}$ signals that this object originates
from short distances. This estimate leads to 
\begin{equation}
\frac{P_{\Delta\tau}(k)}{c_{{\rm ren}}^{2}(k_{\rm{ren}},z)k^{2}P_{{\rm 11}}(k,z)}\sim\left(\frac{k}{k_{{\rm NL}}}\right)^{2-n}
\end{equation}
where $n\approx-1.5$ parameterizes the $k$ scaling of the dimensionful
power spectrum near $k\sim0.1h$/Mpc. Since on $k/k_{\mathrm{NL}}\sim O(0.1)$,
the suppression is expected to be $O(10^{-3})$ which allows us to
drop this term from further analysis.

\subsection{\label{subsec:peculiar}Peculiar counterterm property associated with the large spectral
index}

Interestingly within the literature, one often takes the integration
cutoff $\Lambda\rightarrow\infty$ since the $P_{13}$ integral is
finite for pure adiabatic ICs and the associated parameter $c^{2}=c_{\Lambda}^{2}+c_{{\rm ren}}^{2}$
is also finite \citep{Carrasco_2012,Hertzberg_2014}. Accordingly,
the often quoted value of sound speed $c^{2}(z=0;\Lambda\rightarrow\infty)\sim O(1)\;{\rm \left(Mpc/h\right)^{2}>0}$
is a bare value which is directly measured from N-body simulations/data
\citep{Carrasco_2012,Carrasco:2013sva,Hertzberg_2014,Foreman:2015lca,Baldauf_2015,Baldauf:2015tla,Angulo:2015eqa,Foreman:2015uva}.
However, this approach is unsuitable when studying a wide range of
mixed blue-isocurvature ICs with two additional free model parameters,
$n_{{\rm iso}}$ and $\alpha$. The UV part of $P_{13}$ integral
which can be written as 
\begin{equation}
P_{13}^{{\rm UV}}(k,\Lambda)\approx-k^{2}P_{{\rm 11}}(k)\frac{61}{630\pi^{2}}\int^{\Lambda}dqP_{{\rm 11}}(q)\label{eq:P13UV}
\end{equation}
is divergent for primordial isocurvature spectra with $n_{{\rm iso}}\gtrsim3$.
Since this term is combined with the counterterm Eq.~(\ref{eq:Pctr})
containing the $c^{2}$ coefficient to produce a finite term, $c^{2}$
does not converge as $\Lambda\rightarrow\infty$. Although one can
still measure the bare parameter $c^{2}$ directly from N-body simulations,
a direct comparison with the adiabatic scenario is not particularly
meaningful given that the parameters in the two cases have different
interpretations. For instance unlike the adiabatic case, if $n_{{\rm iso}}=3.75$
and $\alpha=0.25$, the bare value $c^{2}\left({\rm Mpc/h}\right)^{-2}$
has a magnitude of $O(-20)$ as $\Lambda$ goes to $O(100{\rm h/Mpc})$ at a redshift of $z=0$. In Fig.~\ref{fig:eft-fastpm-compare} we plotted EFT curves obtained for mixed ICs with negative value of fitted $c^2$ parameter at different redshifts.
Although this negative value may naively seem alarming, this coefficient
merely represents a coefficient of an EFT operator consistent with
symmetries and power counting, and we did not make assumptions of
positivity of this coefficient.

\begin{figure}
\begin{centering}
\includegraphics[scale=0.5]{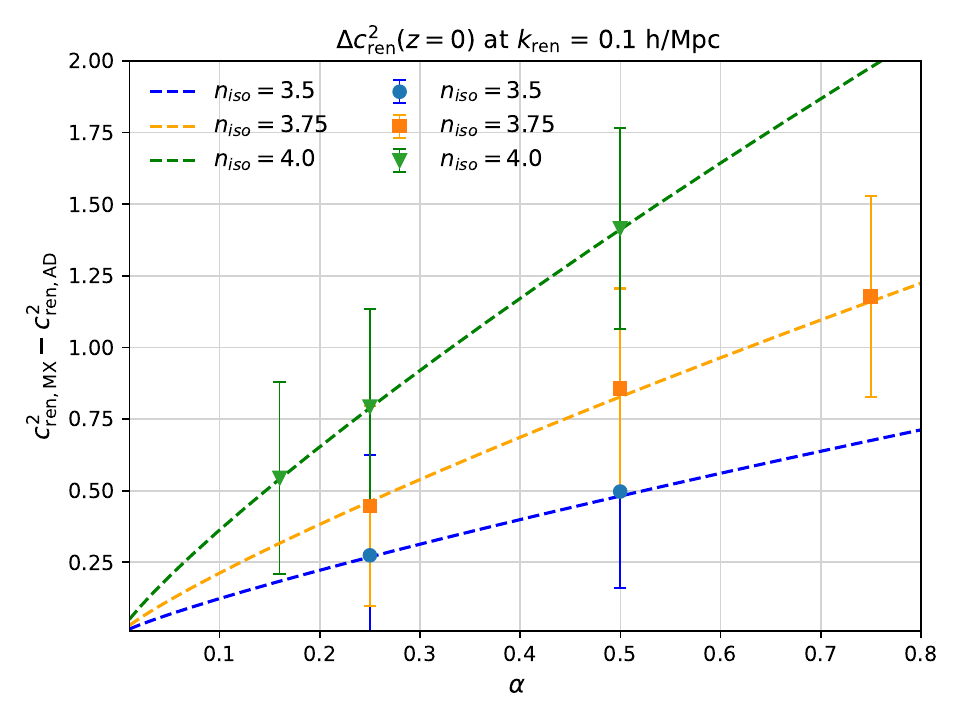}\includegraphics[scale=0.5]{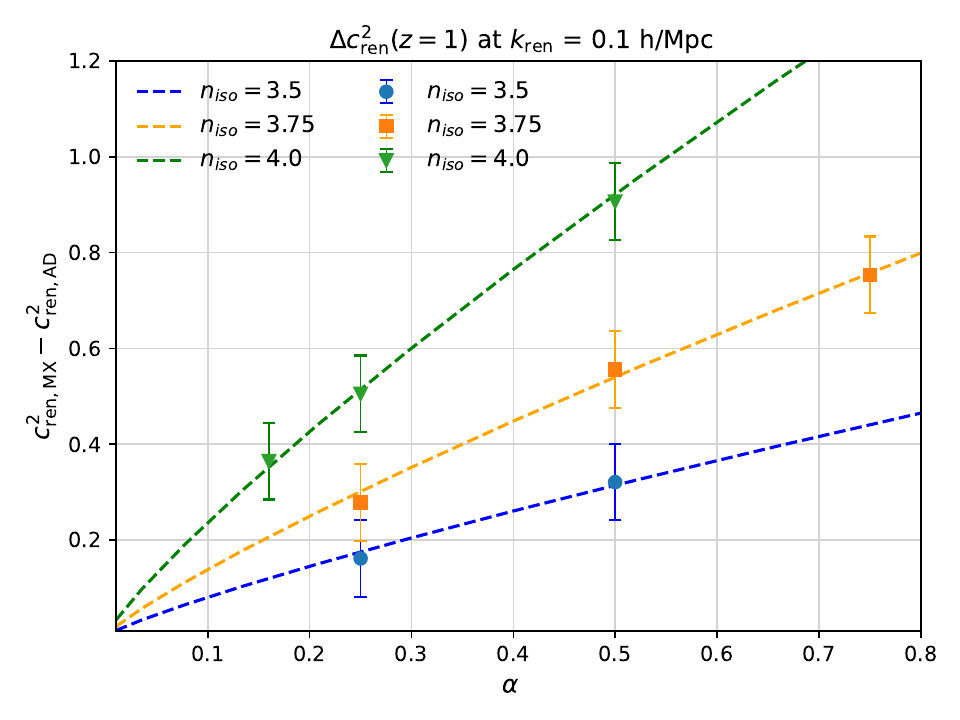}
\par\end{centering}
\caption{\label{fig:delta_c2ren}Plot showing the difference in the renormalized
value of the effective sound speed squared $c_{{\rm ren}}^{2}$ between
mixed and pure ICs obtained by fitting Eq.~(\ref{eq:renormalized_1loopPmm})
to the numerical data from N-body simulations at $z=0$ and $1$.
Additionally, we also plot (dashed curves) a semi-analytical empirical fitting
formula derived in Appendix \ref{sec:Semi-analytic-expression-for-csrenMX}.
Note that the estimation of $c_{{\rm ren}}^{2}$ from N-body data
at higher redshifts has smaller error bars since the number of modes
available for fitting (with smaller cosmic variance and theoretical
error) increases with $z$. Note that the error bars in this figure are given by cosmic variance and set an upper limit on the true statistical error. We have used phase-matched simulations to determine $c^2$, which means that the AD and MX simulations have the same random seeds. This procedure cancels cosmic variance (see e.g. \cite{Villaescusa-Navarro:2018bpd} for a discussion) such that the true statistical error bars should be significantly smaller.}
\end{figure}

More importantly, we note that the only IR-surviving quantity inherited
from the UV effects is the renormalized parameter $c_{{\rm ren}}^{2}$.
Hence, we compare the renormalized parameter $c_{{\rm ren}}^{2}(k_{\rm{ren}},z)$
(instead of the bare parameter) between different cosmological scenarios
while keeping $\Lambda\gg O(10)k_{\mathrm{NL}}$ arbitrary.\footnote{Subleading terms of $O\left(k/\Lambda\right)^{n}$ for $n\geq2$ can
become important for higher order loop calculations for a finite value
of $\Lambda$. These terms can be neglected only in the limit where
$k/\Lambda\rightarrow0$ for modes of interest $k\ll\Lambda$. In
\citep{Carrasco:2013sva} the authors show that for two-loop evaluations
subleading finite terms of $O\left(k/\Lambda\right)^{n}$ must be
included for $\Lambda\lesssim O(20)k_{{\rm NL}}$.} In Fig.~\ref{fig:delta_c2ren} we plot the difference in the value
of renormalized sound speed squared parameter between mixed and pure
adiabatic ICs obtained at a renormalization scale $k_{{\rm ren}}=0.1\rm{h/Mpc}$
by matching the \eft one-loop perturbation theory to the N-body simulations.\footnote{Here we choose the subtraction scheme with a scale $k_{{\rm ren}}$ such that
\begin{equation}
c_{{\rm ren,AD}}^{2}(k_{{\rm ren}},z)=c_{{\rm AD}}^{2}(\Lambda,z)-\frac{P_{13,{\rm AD}}(k_{{\rm ren}},\Lambda,z)}{2k_{{\rm ren}}^{2}P_{{\rm 11, {\rm AD}}}(k_{\rm ren},z)}
\end{equation}
and similarly for the mixed initical condition case.
} 
The renormalized value $c_{{\rm ren,AD}}^{2}\left(k_{{\rm ren}}=0.1{\rm h/Mpc}\right)$
at $z=0$ for pure adiabatic ICs is $\approx14{\rm \left(Mpc/h\right)^{2}}\sim1/k_{{\rm NL}}^{2}$
for $k_{{\rm NL}}\approx0.3{\rm h/Mpc}$. Compared to a large difference
between the bare $c^{2}$ parameters, the difference, $\Delta c_{{\rm ren}}^{2}=c_{{\rm ren,MX}}^{2}-c_{{\rm ren,AD}}^{2}$,
of the renormalized value for the two ICs is $\approx\left(5-10\right)\%$
for high blue spectral indices and with a sizable isocurvature fraction,
$\alpha\sim O(0.1)$. In Appendix \ref{sec:Semi-analytic-expression-for-csrenMX}
we discuss and give a semi-analytical empirical fitting function for
$c_{{\rm ren,MX}}^{2}-c_{{\rm ren,AD}}^{2}$ for values of $O(0.01)\lesssim\alpha<1$
and $2\leq n_{{\rm iso}}\leq4$ for redshifts $z\leq2$.

In order to see the effect of a change in the value of renormalized
fluid parameter, we take the ratio
\begin{equation}
\frac{\Delta P_{c_{{\rm ren}}}(k,z)}{P_{11}(k,z)}  =2\Delta c_{{\rm ren}}^{2}(k_{\rm{ren}},z)k^{2}\\
  =\frac{2\Delta c_{{\rm ren}}^{2}(k_{\rm{ren}},z)}{100}\left(\frac{k}{0.1}\right)^{2}
\end{equation}
which says that an $O(1)$ change in $\Delta c_{{\rm ren}}^{2}$ can
result in a $>1\%$ deviation in the non-linear matter power spectrum
at $k>0.1{\rm h/Mpc}$. This is consistent with the results of Fig.~\ref{fig:delta_c2ren}.
Therefore, when requiring high precision theoretical perturbative
matter power spectrum for mixed ICs with $\alpha\sim O(0.1)$, we
propose using the semi-analytic empirical fitting function for high
blue spectral indices. However, for small isocurvature amplitudes,
$\alpha\lesssim O(0.01)$, it is sufficient to take $c_{{\rm ren,MX}}^{2}\approx c_{{\rm ren,AD}}^{2}$.
Additionally, when working with the biased tracers like galaxy/halo
power spectrum, we note that the one-loop renormalized fluid parameter
$c_{{\rm ren}}^{2}$ is degenerate with the $b_{\nabla^{2}\delta}$
bias coefficient, as we will discuss more below. In forecasts of experiments
that do not break this degeneracy, a precise value of $c_{{\rm ren}}^{2}$
is not required since we marginalize over the bias parameters. In
the present work, it will thus be adequate to work with the renormalized
EFT or bias parameters of adiabatic ICs and apply them directly to
the analysis/forecasts for mixed IC cosmologies.

\section{\label{sec:Galaxy-power-spectrum}Galaxy power spectrum and bispectrum
model}

We use the one-loop galaxy power spectrum and tree-level bispectrum
obtained through a bias expansion for the galaxy density contrast
by including all operators allowed by Galilean symmetry up to cubic
order in magnitude as the coarse-grained linear overdensity $\delta_{\Lambda}^{(1)}$
\citep{Desjacques_2018},
\begin{align}
\delta_{g}(x) & =\sum_{\mathcal{O}}\left(b_{\mathcal{O}}+\epsilon_{\mathcal{O}}(x)\right)\mathcal{O}(x)+b_{\epsilon}\epsilon(x)\\
 & =b_{1}\delta(x)+b_{\epsilon}\epsilon(x)\nonumber \\
 & +\frac{b_{2}}{2}\delta^{2}(x)+b_{\mathcal{G}_{2}}\mathcal{G}_{2}(x)+\epsilon_{\delta}(x)\delta(x)\nonumber \\
 & +b_{\delta\mathcal{G}_{2}}\delta(x)\mathcal{G}_{2}(x)+\frac{b_{3}}{6}\delta^{3}(x)+b_{\mathcal{G}_{3}}\mathcal{G}_{3}(x)+b_{\Gamma_{3}}\Gamma_{3}(x)+\epsilon_{\delta^{2}}(x)\delta^{2}(x)+\epsilon_{\mathcal{G}_{2}}(x)\mathcal{G}_{2}(x)\nonumber \\
 & +b_{\nabla^{2}\delta}\nabla^{2}\delta(x)+b_{\nabla^{2}\epsilon}\nabla^{2}\epsilon(x)\label{eq:galaxy_bias_expn}
\end{align}
where all the operators $\mathcal{O}$ in the above expression are
considered to be coarse-grained and the subscript $\Lambda$ is dropped
for brevity. In Fourier space the Laplacian takes the form $\nabla^{2}\rightarrow\left(k/k_{*}\right)^{2}$
where $k_{*}$ is some characteristic scale of clustering for biased tracers
and we restrict to scales $k/k_{*}\ll1$. Hence every insertion of
a Laplacian is equivalent to a second order correction to an operator
$\mathcal{O}$ and the derivative operators in the last-line of Eq.~(\ref{eq:galaxy_bias_expn})
are counted approximately as cubic order in bias expansion. Therefore,
Eq.~(\ref{eq:galaxy_bias_expn}) is a double expansion in density
fluctuations and their derivatives. The remaining operator set $\{\delta^{2},\mathcal{G}_{2},\epsilon_{\delta}\delta\}$
and $\{\mathcal{G}_{2}\delta,\delta^{3},\mathcal{G}_{3},\Gamma_{3},\epsilon_{\delta^{2}}\delta^{2},\epsilon_{\mathcal{G}_{2}}\mathcal{G}_{2}\}$
are second and third order respectively where we refer the readers
to \citep{Assassi:2014fva,Desjacques_2018} for definition and details
regarding these operators.

Notably \citep{Chan_2012} and \citep{Assassi:2014fva} have shown
that operators non-local in $\delta$ such as $\mathcal{G}_{2}=\left(\nabla_{i}\nabla_{j}\Phi\right)^{2}-\left(\nabla^{2}\Phi\right)^{2}$
arise naturally due to gravitational evolution and renormalization
requirements respectively. Meanwhile, $\epsilon$ in Eq.~(\ref{eq:galaxy_bias_expn})
is the stochastic noise contribution to the galaxy formation where
we remark that the stochastic fields $\epsilon_{{\rm m}}$, $\epsilon$
and $\epsilon_{\mathcal{O}}$ (here the stochastic operators $\epsilon_{\mathcal{O}}$
are sourced through gravitational evolution) are modeled such that
they are sourced by small-scale fluctuations and are assumed to be
uncorrelated with the long-wavelength perturbative fluctuations at
large separations.

One can choose the renormalization condition (see \citep{Assassi:2014fva}\textbf{
}and \citep{Desjacques_2018} for details regarding the field-redefinition
procedure which we summarize in Appendix \ref{sec:Renormalization-scheme})
such that the one-loop galaxy power spectrum at $O\left(\left(\delta^{(1)}\right)^{4}\right)$
can be written as 
\begin{align}
P_{gg}(k) & =b_{1}^{2}P_{{\rm NL}}(k)+P_{gg}^{{\rm NLO}}(k)+P_{gg,\nabla^{2}\delta}(k)+P_{gg,\epsilon}(k)\label{eq:Pgg}
\end{align}
where the non-linear matter power spectrum up to one-loop in the \eft perturbation theory is given as
\begin{equation}
P_{{\rm NL}}(k)=P_{11}(k)+P_{22}(k)+P_{13}(k)-2c^{2}k^{2}P_{11}(k)
\end{equation}
 and the next-to-leading order one-loop contributions (without stochastic
terms) are given by 
\begin{align}
P_{gg}^{{\rm NLO}}(k) & =b_{1}\left(b_{2}\mathcal{I}_{\delta^{(2)}\delta^{2}}(k)+2b_{\mathcal{G}_{2}}\mathcal{I}_{\delta^{(2)}\mathcal{G}_{2}}(k)+\left(2b_{\mathcal{G}_{2}}+\frac{4}{5}b_{\Gamma_{3}}\right)\mathcal{F}_{\mathcal{G}_{2}}(k)(k)\right)\nonumber \\
 & +b_{2}b_{\mathcal{G}_{2}}\mathcal{I}_{\delta^{2}\mathcal{G}_{2}}(k)+\frac{1}{4}b_{2}^{2}\mathcal{I}_{\delta^{2}\delta^{2}}(k)+b_{\mathcal{G}_{2}}^{2}\mathcal{I}_{\mathcal{G}_{2}\mathcal{G}_{2}}(k)\label{eq:PNLO}
\end{align}
where the loop contributions $\mathcal{I}_{\mathcal{O},\mathcal{O}'}(k)$
are given in Appendix $\ref{sec:Galaxy-one-loop-contributions}$.
The leading derivative contribution is
\begin{equation}
P_{gg,\nabla^{2}\delta}(k)=-2b_{1}b_{\nabla^{2}\delta}\left(\frac{k}{k_{*}}\right)^{2}P_{{\rm 11}}(k)\label{eq:P_lap}
\end{equation}
and the stochastic noise contribution is given as
\begin{equation}
P_{gg,\epsilon}(k)=P_{\epsilon\epsilon}b_{\epsilon}\left(1+2b_{\nabla^{2}\epsilon}\left(\frac{k}{k_{{\rm *}}}\right)^{2}\right).\label{eq:ggeps}
\end{equation}
Note that all these quantities were computed after renormalizing the
composite operators such as $\delta^{2}$ with the renormalization
scheme explained in Appendix~\ref{sec:Renormalization-scheme}. Contributions
from other operators like $\delta^{3},\mathcal{G}_{3},\delta\mathcal{G}_{2}$
do not appear as they are eliminated in the renormalization scheme
used in this paper (see Appendix \ref{sec:Renormalization-scheme}).
The bias term associated with the parameter $b_{\nabla^{2}\epsilon}$
represents a scale-dependent deviation from a purely Poissonian shot
noise (i.e.~the $k^{2}$ dependence makes this scale dependent instead
of being a constant), and represents a (small) correlation of the
stochastic bias on large scales. Depending upon the value of $b_{\nabla^{2}\epsilon}$
it can either enhance or reduce the total power spectrum \citep{Eggemeier_2020,Eggemeier_2021}.
As we will discuss more in depth later, the scale $k_{*}$ in Eq.~(\ref{eq:ggeps})
is close to the nonlinear scale $k_{{\rm NL}}.$ In Sec.~\ref{sec:UV-div-and-corr}
we will note that the higher-derivative stochastic bias term is also
motivated by the renormalization requirements for the next-to-leading
order (NLO) terms.

Due to the assumed Gaussian initial conditions of $\delta_{k_{1}}^{(1)}$,
the galaxy bispectrum at the perturbative order $O\left(\left(\delta^{(1)}\right)^{4}\right)$
is given by the following terms \citep{Scoccimarro_2004}:
\begin{align}
B_{gg}\left(\vec{k}_{1},\vec{k}_{2},\vec{k}_{3}\right) & \sim b_{1}^{3}\left\langle \delta_{k_{1}}^{(1)}\delta_{k_{2}}^{(1)}\delta_{k_{3}}^{(2)}\right\rangle \nonumber \\
 & +b_{1}^{2}b_{2}\left\langle \delta_{k_{1}}^{(1)}\delta_{k_{2}}^{(1)}\delta_{k_{3}}^{2}\right\rangle +b_{1}^{2}b_{\mathcal{G}_{2}}\left\langle \delta_{k_{1}}^{(1)}\delta_{k_{2}}^{(1)}\mathcal{G}_{2k_{3}}\right\rangle \nonumber \\
 & +b_{1}^{2}\left\langle \delta_{k_{1}}\delta_{k_{2}}\left(\epsilon_{\delta}\delta\right)_{k_{3}}\right\rangle +\left\langle \epsilon_{k_{1}}\epsilon_{k_{2}}\left(\epsilon_{\delta}\delta\right)_{k_{3}}\right\rangle +b_{1}\left\langle \epsilon_{k_{1}}\delta_{k_{2}}\left(\epsilon_{\delta}\delta\right)_{k_{3}}\right\rangle +\left\langle \epsilon_{k_{1}}\epsilon_{k_{2}}\epsilon_{k_{3}}\right\rangle +\mbox{permutations }
\end{align}
where the terms in the last line give a mixed stochastic noise contribution
and the vectors $\vec{k}_{1,2,3}$ must form a closed triangle such
that $\vec{k}_{1}+\vec{k}_{2}+\vec{k}_{3}=0.$ Note that the stochastic
contributions represent $k>k_{\mathrm{NL}}$ contributions to the
perturbations which generically will be non-Gaussian. Evaluating the
three-point correlation functions we obtain
\begin{align}
B_{gg}\left(\vec{k}_{1},\vec{k}_{2},\vec{k}_{3}\right) & =\left(F_{2}^{b}\left(\vec{k}_{1},\vec{k}_{2}\right)b_{1}^{2}P_{11}(k_{1})P_{11}(k_{2})+\mbox{cyclic}\right)\nonumber \\
 & +P_{{\rm shot}}\sum_{i=1}^{3}b_{1}^{2}P_{11}(k_{i})+B_{{\rm shot}}\label{eq:bispectrum_Bgg}
\end{align}
where
\begin{equation}
F_{2}^{b}\left(\vec{k}_{1},\vec{k}_{2}\right)=\frac{b_{2}}{2}+b_{\mathcal{G}_{2}}\left(\mu_{12}^{2}-1\right)+b_{1}\left(\frac{5}{7}+\frac{1}{2}\mu_{12}\left(\frac{k_{1}}{k_{2}}+\frac{k_{2}}{k_{1}}\right)+\frac{2}{7}\mu_{12}^{2}\right)
\end{equation}
for $\mu_{12}=\vec{k}_{1}\cdot\vec{k}_{2}/\left(k_{1}k_{2}\right)$,
and $P_{{\rm shot}}$ and $B_{{\rm shot}}$ are the shot noise contributions
whose fiducial values will be discussed in Sec.~\ref{subsec:Fiducial-parameter-values}. At the perturbative
order $\left(\delta^{(1)}\right)^{4}$, the galaxy bispectrum in Eq.~(\ref{eq:bispectrum_Bgg})
is given only by tree-level terms since unlike the power spectrum
in Eq.~(\ref{eq:Pgg}) it does not contain any loop corrections.
Also we point out that the only relevant bias terms at this order
for bispectrum are $b_{1},b_{2}\mbox{, and }b_{\mathcal{G}_{2}}$.

For the theoretical error estimates we consider the following explicit
two-loop error envelope $E_{{\rm gg}}(k,z)$ for the galaxy-galaxy
power spectrum as given in \citep{Chudaykin:2019ock}:
\begin{equation}
E_{{\rm gg}}^{{\rm p}}(k,z)=\left(D_{+}(z)/D_{+}(0)\right)^{4}P_{{\rm gg}}(k,z)\left(\frac{k}{0.45\,{\rm hMpc^{-1}}}\right)^{3.3}\label{eq:error}
\end{equation}
where $P_{{\rm gg}}$ is the galaxy auto-correlation power spectrum
at one-loop order without the stochastic component. We would like to emphasize that the error envelope provided in Eq.~\ref{eq:error} is a conservative estimate of the actual error, typically by a factor of $\sim2$ compared to DM simulations for the majority of the relevant k-values. However, given that this estimate is derived from matching with DM-only N-body simulations and considering the possibility of larger errors in the case of biased tracers, we find it appropriate to still apply this estimate to the biased tracer measurements.\footnote{We found that normalizing the error envelope in Eq.~(\ref{eq:error}) to match the N-body data leads to improvements in the marginalized results by approximately 20-30$\%$.} Overall, a more accurate estimation of the theoretical error specifically for the biased tracers holds the potential for better forecast and fitting results in future analyses. In this regard, we refer the readers to \cite{DAmico:2019fhj}, where the authors conducted a systematic analysis to estimate the theory errors from Mock galaxy samples.
We also remark that we have verified the above error envelope for mixed isocurvature scenarios
and found it to be in good agreement (see Fig. \ref{fig:eft-fastpm-compare}). This can be partially explained
by the smallness of the isocurvature contribution $\alpha(f_{c}/3)^{2}\ll1$
(see Eq.~(\ref{eq:Pm_analytical})) allowed by the existing data
in the $k$-range that the present forecast experiments are sensitive
to and the stability of the error envelope to small perturbations.
Similarly, for the tree-level bispectrum we approximate the error
envelope with
\begin{equation}
E_{{\rm gg}}^{{\rm b}}(k,z)=3\left(D_{+}(z)/D_{+}(0)\right)^{2}B_{{\rm gg}}(k,z)\left(\frac{k_{t}/3}{0.31\,{\rm hMpc^{-1}}}\right)^{1.8}\label{eq:bispecenv}
\end{equation}
where $B_{{\rm gg}}(k,z)$ is now the tree-level bispectrum (without
$B_{{\rm shot}}$) evaluated using Eq.~(\ref{eq:bispectrum_Bgg})
and $k_{{\rm t}}=\left|\vec{k}_{{\rm 1}}+\vec{k}_{{\rm 2}}+\vec{k}_{{\rm 3}}\right|$. 

\section{\label{sec:UV-div-and-corr}UV divergences and correction}

Similar to the discussion in Sec.~\ref{sec:Review-of-eft} where
we reviewed the need to renormalize the loop integrals in the matter
power spectrum, the next to leading order $P_{gg}^{{\rm NLO}}$ terms
(Eq.~(\ref{eq:PNLO})) in the galaxy two-point correlation function
may also show similar $\Lambda$-dependence and UV divergences. Specifically,
the one-loop terms in Eq.~(\ref{eq:PNLO}) have the following UV limits
\begin{align}
\lim_{k\rightarrow0}\mathcal{I}_{\delta^{(2)}\delta^{2}}(k) & =\frac{-1}{42\pi^{2}}k^{2}\int^{\Lambda}dq\left(P_{11}(q)\right)^{2}+O(k^{4}),\nonumber \\
\lim_{k\rightarrow0}\mathcal{I}_{\delta^{(2)}\mathcal{G}_{2}}(k) & =\frac{-1}{21\pi^{2}}k^{4}\int^{\Lambda}dq\left(\frac{P_{11}(q)}{q}\right)^{2}+O(k^{6}),\nonumber \\
\lim_{k\rightarrow0}\mathcal{F}_{\mathcal{G}_{2}}(k) & =-\frac{16}{21\pi^{2}}k^{2}P_{11}(k)\int^{\Lambda}dqP_{11}(q)+O(k^{4}),\nonumber \\
\lim_{k\rightarrow0}\mathcal{I}_{\delta^{2}\delta^{2}}(k) & =\frac{1}{2\pi}\int^{\Lambda}dq\left(qP_{11}(q)\right)^{2}+\frac{k^{2}}{4\pi^{2}}\int^{\Lambda}dqP_{11}(q)\left(\frac{4}{3}q\partial_{q}P_{11}(q)+\frac{2}{3}q^{2}\partial_{q}^{2}P_{11}(q)\right)+O(k^{4}),\nonumber \\
\lim_{k\rightarrow0}\mathcal{I}_{\mathcal{G}_{2}\mathcal{G}_{2}}(k) & =\frac{8}{15\pi^{2}}k^{4}\int^{\Lambda}dq\left(\frac{P_{11}(q)}{q}\right)^{2}+O(k^{6}),\nonumber \\
\lim_{k\rightarrow0}\mathcal{I}_{\delta^{2}\mathcal{G}_{2}}(k) & =\frac{-2}{3\pi^{2}}k^{2}\int^{\Lambda}dq\left(P_{11}(q)\right)^{2}+O(k^{4}).\label{eq:UV-PNLO}
\end{align}
 In the limit $k/\Lambda\rightarrow0$, we expect the galaxy power
spectrum to be well-defined by the leading perturbative term associated
with the linear bias and a stochastic field $\epsilon$. Hence all
higher order bias terms should become insignificant at very large
scales. One notes from Eq.~set~(\ref{eq:UV-PNLO}) that the leading
$O(k^{0})$ contribution from $\mathcal{I}_{\delta^{2}\delta^{2}}(k)$
term does not vanish as $k/\Lambda\rightarrow0$. Since it is clear
that any $O(k^{0})$ quantity will be degenerate with the stochastic
contribution $b_{\epsilon}^{2}\left\langle \epsilon_{k}\epsilon_{p}\right\rangle $,
we remove the $\Lambda$-dependent $O(k^{0})$ diverging contribution
from $\mathcal{I}_{\delta^{2}\delta^{2}}(k)$ by subtracting the $k/\Lambda\rightarrow0$
leading term $\frac{1}{2\pi}\int^{\Lambda}dq\left(qP_{11}(q)\right)^{2}$
from $\mathcal{I}_{\delta^{2}\delta^{2}}(k)$ and redefining the stochastic
bias parameter $b_{\epsilon}$ in Eq.~(\ref{eq:ggeps}) to absorb
this contribution. Hence, the first renormalized $\mathcal{I}_{\delta^{2}\delta^{2}}$
is written as 
\begin{equation}
\mathcal{I}_{\delta^{2}\delta^{2}}^{{\rm [R1]}}(k)\equiv\mathcal{I}_{\delta^{2}\delta^{2}}(k)-\frac{1}{2\pi}\int_{0}^{\Lambda}dq\left(qP_{11}(q)\right)^{2}=\frac{k^{2}}{4\pi^{2}}\int_{0}^{\Lambda}dqP_{11}(q)\left(\frac{4}{3}q\partial_{q}P_{11}(q)+\frac{2}{3}q^{2}\partial_{q}^{2}P_{11}(q)\right)+O(k^{4}).\label{eq:correctedId2d2}
\end{equation}
where only $k^{2}$ independent term has been removed and thus not
fully renormalized. This freedom to absorb the contact terms by explicitly
reparameterizing the stochastic bias parameters was first emphasized
in \citep{PhysRevD.74.103512} and also noted in \citep{Senatore:2014eva,Assassi:2014fva}.
After this procedure, we find that all one-loop contributions in $P_{gg}^{{\rm NLO}}$
scale as either $k^{2}$ or $k^{4}$ in the limit $k\rightarrow0$.
For the general pure adiabatic ICs, the correction made above to the
$\mathcal{I}_{\delta^{2}\delta^{2}}(k)$ one-loop term is sufficient
to ensure that all the one-loop integrals in Eq.~(\ref{eq:PNLO})
converge and thus vanish in the limit $k\rightarrow0$. One can see
this explicitly in \citep{Simonovic:2017mhp} which gives a range
of spectral index-related parameters $\nu\in\left(\nu_{\mbox{IR}},\nu_{\mbox{UV}}\right)$
for which the integrals converge. For pure adiabatic ICs in $\Lambda{\rm CDM}$
characterized by near-scale-invariance, it is easy to check that the
one-loop integrals in Eq.~set~(\ref{eq:UV-PNLO}) are convergent
as $q\rightarrow\infty$, and the terms vanish in the limit $k\rightarrow0$
due to an overall $k^{n}$ scaling. However, we note that similar
to $P_{13}$ loop term, the integral $\mathcal{F}_{\mathcal{G}_{2}}(k)$
has a finite non-negligible support from non-linear scales $q\sim q_{{\rm NL}}$
and thus includes contribution from scales where our perturbation
theory is known to be inaccurate. This naively worrisome inaccuracy
however does not contribute to the physical observables as their contributions
are absorbed by the renormalization scheme used for $b_{\nabla^{2}\delta}$.

Unlike the case for pure adiabatic ICs, the UV sensitivity for mixed
ICs with large blue spectral indices is more severe since the one-loop
terms $\mathcal{I}_{\delta^{(2)}\delta^{2}}(k)$, $\mathcal{I}_{\delta^{2}\delta^{2}}^{{\rm [R1]}}(k)$
and $\mathcal{I}_{\delta^{2}\mathcal{G}_{2}}(k)$ are UV divergent
when $n_{{\rm iso}}\gtrsim3.5$ and $\mathcal{F}_{\mathcal{G}_{2}}(k)$
is divergent when $n_{{\rm iso}}\gtrsim3$. We note that the leading
UV behavior of $\mathcal{I}_{\delta^{(2)}\delta^{2}}(k)$, $\mathcal{I}_{\delta^{2}\delta^{2}}^{{\rm [R1]}}(k)$
and $\mathcal{I}_{\delta^{2}\mathcal{G}_{2}}(k)$ correspond to contact
terms of the form $O(k^{n})$ for $n\geq2$ and are absorbed by the
renormalization condition of higher order derivatives of stochastic
terms in our bias expansion. For instance, the bias term $b_{\nabla^{2}\epsilon}$
in Eq.~(\ref{eq:ggeps}) is redefined after absorbing all $O(k^{2}\times\mbox{constant})$
contact terms. This will be part of our renormalization scheme for
the bias expansion. Likewise, the leading divergent piece from $\mathcal{F}_{\mathcal{G}_{2}}(k)$
which scales as $O\left(k^{2}P_{11}(k)\right)$ is removed by redefining
the $b_{\nabla^{2}\delta}$ bias parameter of the leading higher-derivative
operator. These are shape-changing contributions and are often interpreted
as scale-dependent additions to the linear bias $b_{1}$.

Finally, the new renormalized one-loop terms in $P_{gg}^{{\rm NLO}}$
are given as\footnote{In this work, we are mainly interested in blue isocurvature with spectral
indices $n_{{\rm iso}}\leq4$. Hence, the corrections given below
are sufficient to ensure that the one-loop terms are convergent.}
\begin{eqnarray}
\mathcal{I}_{\delta^{(2)}\delta^{2}}^{{\rm [R]}}(k) & =&\mathcal{I}_{\delta^{(2)}\delta^{2}}(k)-\frac{1}{2\pi}\int_{0}^{\Lambda}dq\left(qP_{11}(q)\right)^{2}-C_{\delta^{(2)}\delta^{2}}k^{2},\nonumber \\
\mathcal{I}_{\delta^{2}\delta^{2}}^{{\rm [R]}}(k) & =&\mathcal{I}_{\delta^{2}\delta^{2}}(k)-2\int_{q}P_{{\rm 11}}(q)P_{{\rm 11}}(q)-C_{\delta^{2}\delta^{2}}k^{2},\nonumber \\
\mathcal{I}_{\delta^{2}\mathcal{G}_{2}}^{{\rm [R]}}(k) & =&\mathcal{I}_{\delta^{2}\mathcal{G}_{2}}(k)-C_{\delta^{2}\mathcal{G}_{2}}k^{2},\nonumber \\
\mathcal{F}_{\mathcal{G}_{2}}^{{\rm [R]}}(k) & = &\mathcal{F}_{\mathcal{G}_{2}}(k)-C_{\mathcal{F}_{\mathcal{G}_{2}}}k^{2}P_{11}(k),\nonumber \\
\mathcal{I}_{\delta^{(2)}\mathcal{G}_{2}}^{{\rm [R]}}(k) & = &\mathcal{I}_{\delta^{(2)}\mathcal{G}_{2}}(k),\nonumber \\
\mathcal{I}_{\mathcal{G}_{2}\mathcal{G}_{2}}^{{\rm [R]}}(k) & = &\mathcal{I}_{\mathcal{G}_{2}\mathcal{G}_{2}}(k),\label{eq:renormalized_PNLO}
\end{eqnarray}
where the $\Lambda$-dependent coefficients $C_{\mathcal{O},\mathcal{O}'}$
are given by the following expressions
\begin{align}
C_{\delta^{(2)}\delta^{2}} & \equiv\frac{1}{2}\left.\frac{d^{2}\mathcal{I}_{\delta^{(2)}\delta^{2}}(k)}{dk^{2}}\right|_{k=0}=-\frac{1}{42\pi^{2}}\int_{0}^{\Lambda}dq\left(P_{11}(q)\right)^{2},\label{eq:C1}\\
C_{\delta^{2}\delta^{2}} & \equiv\frac{1}{2}\left.\frac{d^{2}\mathcal{I}_{\delta^{2}\delta^{2}}(k)}{dk^{2}}\right|_{k=0}=\frac{1}{4\pi^{2}}\int_{0}^{\Lambda}dqP_{11}(q)\left(\frac{4}{3}q\partial_{q}P_{11}(q)+\frac{2}{3}q^{2}\partial_{q}^{2}P_{11}(q)\right),\label{eq:C2}\\
C_{\delta^{2}\mathcal{G}_{2}} & \equiv\frac{1}{2}\left.\frac{d^{2}\mathcal{I}_{\delta^{2}\mathcal{G}_{2}}(k)}{dk^{2}}\right|_{k=0}=\frac{-2}{3\pi^{2}}\int_{0}^{\Lambda}dq\left(P_{11}(q)\right)^{2},\label{eq:C3}\\
C_{\mathcal{F}_{\mathcal{G}_{2}}} & \equiv\frac{1}{P_{11}(k)}\frac{1}{2}\left.\frac{d^{2}\mathcal{F}_{\mathcal{G}_{2}}(k)}{dk^{2}}\right|_{k=0}=-\frac{16}{21\pi^{2}}\int_{0}^{\Lambda}dqP_{11}(q).\label{eq:C4}
\end{align}
After subtracting the $C_{\mathcal{O},\mathcal{O}'}$ dependent pieces,
the renormalized one-loop contributions (except $\mathcal{F}_{\mathcal{G}_{2}}^{{\rm [R]}}(k)$)
have the same $O(k^{4}$) behavior on large scales, which are subdominant
to other contributions. The leading $O(k^{0}$) and $O(k^{2})$ contributions
are now carried by the stochastic terms associated with the renormalized
biases $b_{\epsilon}$ and $b_{\nabla^{2}\epsilon}$. Similarly, the
leading $O(k^{2}P_{11}(k))$ contributions are controlled by the EFT
parameter $c_{{\rm ren}}^{2}$ and the Laplacian bias parameter $b_{\nabla^{2}\delta}$,
leading to a parametric degeneracy with respect to the galaxy correlation
function observable at the quartic order in our bias expansion. However,
these degeneracies are broken when bispectrum, redshift-space distortion
(RSD) effects and/or matter and tracer cross-correlation power spectrum
are included in the set of observables. Using the expressions given
in Eq.~set~(\ref{eq:renormalized_PNLO}), it is easy to check that
the power spectra for the pure adiabatic and mixed (adiabatic and
blue isocurvature) ICs are equivalent at large scales ($k\rightarrow0$)
for any arbitrary value of the cutoff scale $\Lambda$. This is what
we expect on long wavelengths where the adiabatic contributions dominate.

Finally, the complete galaxy power spectrum up to one-loop order is
given by the following set of bias parameters 
\begin{equation}
\left\{ b_{1},b_{2},b_{\mathcal{G}_{2}},b_{\Gamma_{3}},b_{\nabla^{2}\delta},b_{\epsilon},b_{\nabla^{2}\epsilon}\right\} 
\end{equation}
and the corresponding one-loop contributions are given in Eq.~set~(\ref{eq:renormalized_PNLO}).
Note that $\{b_{\epsilon},b_{\nabla^{2}\delta},b_{\nabla^{2}\epsilon}\}$
are $\Lambda$-dependent. When we marginalize over these parameters
in the numerical procedure, we are marginalizing over only the finite
pieces that are left over after canceling the $\Lambda$-dependent
pieces. We will call the parameters after the subtraction discussed
above as $\{b_{\epsilon}^{[R]},b_{\nabla^{2}\delta}^{[R]},b_{\nabla^{2}\epsilon}^{[R]}\}$.

For example, the galaxy power spectrum has a contribution
\begin{align*}
P_{gg}(k) & \supset b_{1}\left(2b_{\mathcal{G}_{2}}+\frac{4}{5}b_{\Gamma_{3}}\right)\mathcal{F}_{\mathcal{G}_{2}}(k)-2b_{1}b_{\nabla^{2}\delta}\left(\frac{k}{k_{*}}\right)^{2}P_{{\rm 11}}(k)
\end{align*}
where $\mathcal{F}_{\mathcal{G}_{2}}(k)$ is divergent as $\Lambda/k_{\mathrm{NL}}\rightarrow\infty$.
This divergent piece can be subtracted as
\begin{align}
b_{1}\left(2b_{\mathcal{G}_{2}}+\frac{4}{5}b_{\Gamma_{3}}\right)\mathcal{F}_{\mathcal{G}_{2}}(k)-2b_{1}b_{\nabla^{2}\delta}k^{2}P_{{\rm 11}}(k) & =b_{1}\left(2b_{\mathcal{G}_{2}}+\frac{4}{5}b_{\Gamma_{3}}\right)\left(\mathcal{F}_{\mathcal{G}_{2}}(k)-C_{\mathcal{F}_{\mathcal{G}_{2}}}k^{2}P_{11}(k)\right)\nonumber \\
 & -2b_{1}\left(b_{\nabla^{2}\delta}-\left(b_{\mathcal{G}_{2}}+\frac{2}{5}b_{\Gamma_{3}}\right)C_{\mathcal{F}_{\mathcal{G}_{2}}}\right)k^{2}P_{{\rm 11}}(k)\\
 & =b_{1}\left(2b_{\mathcal{G}_{2}}+\frac{4}{5}b_{\Gamma_{3}}\right)\mathcal{F}_{\mathcal{G}_{2}}^{[{\rm R}]}(k)-2b_{1}b_{\nabla^{2}\delta}^{[{\rm R}]}k^{2}P_{{\rm 11}}(k)
\end{align}
where $C_{\mathcal{F}_{\mathcal{G}_{2}}}$ is defined in Eq.~(\ref{eq:C4}),
and we defined the renormalized Laplacian bias as
\begin{equation}
b_{\nabla^{2}\delta}^{[{\rm R}]}=b_{\nabla^{2}\delta}-\left(b_{\mathcal{G}_{2}}+\frac{2}{5}b_{\Gamma_{3}}\right)C_{\mathcal{F}_{\mathcal{G}_{2}}}.
\end{equation}
It is this type of renormalized parameter such as $b_{\nabla^{2}\delta}^{[{\rm R}]}$
that will be marginalized over in the forecast. As mentioned earlier
in Sec.~\ref{sec:Review-of-eft}, we will evaluate the renormalized
counterterm and bias parameters for adiabatic ICs and use them directly
for forecasts related to mixed ICs.

\section{\label{sec:Parameter-set-and}Parameter set and fiducial values}

In this section, we present the relevant experimental characteristics
for Euclid \citep{Euclid} and MegaMapper \citep{Schlegel:2019eqc}
that we use in our forecast. For a detailed discussion on these two
and several other futuristic high-redshifts surveys, we refer the
reader to \citep{Sailer:2021yzm}. We also list the fiducial parameter
values, errors used, marginalized parameters, and other assumptions
made in the Fisher forecast.

\subsection{\label{subsec:Surveys}Surveys}

The Euclid satellite (planned for launch in 2023) will measure star-forming
luminous galaxies containing ${\rm H\alpha}$ emitters using a near-infrared
telescope \citep{Euclid,Euclid:2019clj}. The satellite is expected
to map nearly 15,000 square degrees of sky probing a redshift range
up to $z\sim2$. In order to model the expected population density
of ${\rm H\alpha}$ emitters at high-redshifts, we consider its redshift
distribution per solid-angle $dN(z)/d\Omega dz$ by averaging the
data given in \citep{Pozzetti:2016cch} and \citep{Euclid:2019clj}.
Integrating this distribution over the surveyed area and a redshift
bin yields $N(\bar{z})$, the expected number of galaxies to be detected.

Similar to the analysis in \citep{Chudaykin:2019ock}, we will consider
6 non-overlapping redshift bins with the corresponding mean galaxy
number densities $n(\bar{z})=N(\bar{z})/V(\bar{z})$ provided in Table
\ref{tab:Specification-of-the}. We evaluate the comoving volume $V(\bar{z})$
of a redshift bin centered around $\bar{z}$ using the expression
\begin{equation}
V(\bar{z})=\frac{4\pi}{3}\left(r_{\bar{z}+\Delta z/2}^{3}-r_{\bar{z}-\Delta z/2}^{3}\right)
\end{equation}
where $r_{z}$ is the comoving distance up to a redshift of $z$.
Note that the surveyed volume $V_{s}(z)=V(z)f_{{\rm sky}}$ where
we will set the fraction of sky surveyed $f_{{\rm sky}}\approx0.35$
for both Euclid and MegaMapper. For each redshift bin centered at $\bar{z}$, we set the fiducial linear bias parameter for Euclid surveyed galaxy samples as $b_{1}=\sqrt{1+\bar{z}}$ \citep{Euclid:2021qvm}.

\begin{table}[H]
\begin{centering}
\begin{tabular}{|c|c|c|c|c|c|c|}
\hline 
$\bar{z}$ & 0.8 & 1.0 & 1.2 & 1.4 & 1.6 & 1.8\tabularnewline
\hline 
\hline 
$n(\bar{z})$[$10^{-3}{\rm h^{3}Mpc^{-3}}]$ & 2.08 & 1.18 & 0.7 & 0.42 & 0.26 & 0.19\tabularnewline
\hline 
\end{tabular}
\par\end{centering}
\caption{\label{tab:Specification-of-the}Mean galaxy number density $n(\bar{z})=N(\bar{z})/V(\bar{z})$
expected to be measured by the Euclid survey at a given redshift bin centered at $\bar{z}$
for 6 non-overlapping redshift bins of width $\Delta z=0.2$.}
\end{table}

\begin{table}[H]
\begin{centering}
\begin{tabular}{|c|c|c|c|c|}
\hline 
$\bar{z}$ & 2.0 & 3.0 & 4.0 & 5.0\tabularnewline
\hline 
\hline 
$n(\bar{z})$$[10^{-3}{\rm h^{3}Mpc^{-3}}]$ & 0.98 & 0.12 & 0.1 & 0.04\tabularnewline
\hline 
\end{tabular}
\par\end{centering}
\caption{\label{tab:MM}`Fiducial' mean galaxy number density $n(\bar{z})=N(\bar{z})/V(\bar{z})$
expected to be measured by MM at a given redshift bin centered at $\bar{z}$
for 4 non-overlapping redshift bins of width $\Delta z=1.0$.}
\end{table}

While
the Euclid survey is under construction and expected to be launched
in 2023 providing us with a plethora of cosmological data for following
6 years, we will also give a forecast of a futuristic experiment,
MegaMapper (MM) \citep{Schlegel:2019eqc}, that is still at the design
stage. Similar to Euclid, MM will cover nearly 14,000 square degrees
of the sky but surveying galaxies at higher redshift than Euclid ($2<z<5$)
during an observation period of roughly 5 years. MM will primarily
look for Lyman Break Galaxies (LBGs) to distinguish the higher redshift
galaxies\footnote{The technique of observing LBGs seems to favor $z\gtrsim3$ in terms
of UV light characteristics and telescope technology for the UV selection
method \citep{doi:10.1146/annurev-astro-081710-102542}.}. For the forecast, we will consider the fiducial MM experiment consisting
of $\approx40$ million galaxies in the aforementioned redshift range for
which the number density and redshift bins are listed in Table 2 of
\citep{Ferraro:2019uce} and also given in Tab.~(\ref{tab:MM}) here. We choose this fiducial case instead of
the idealized case (see Table 1 of \citep{Ferraro:2019uce}) to be
more conservative. These fiducial functions give a good fit to previous
observations on large scales, where we expect the $b_{1}$ parameter
to dominate. Note that the galaxy biases of Tables 1 and 2 of \citep{Ferraro:2019uce}
are magnitude dependent, in contrast with the bias parameters that
would map smooth density fields to all galaxy counts. This feature
is one of the contributing factors to the definition of fiducial and
idealized. It is also interesting to note that Table 2 of \citep{Ferraro:2019uce}
has a non-monotonic behavior of $b_{1}(z)$ as it has a local maximum
at $z\approx3$. This feature seems to be partially responsible for
features that we see in the redshift dependencies of various sensitivity
plots in Sec.~\ref{sec:Results-and-Discussion}. 
\begin{figure}[ht]
\begin{centering}
\includegraphics[scale=0.45]{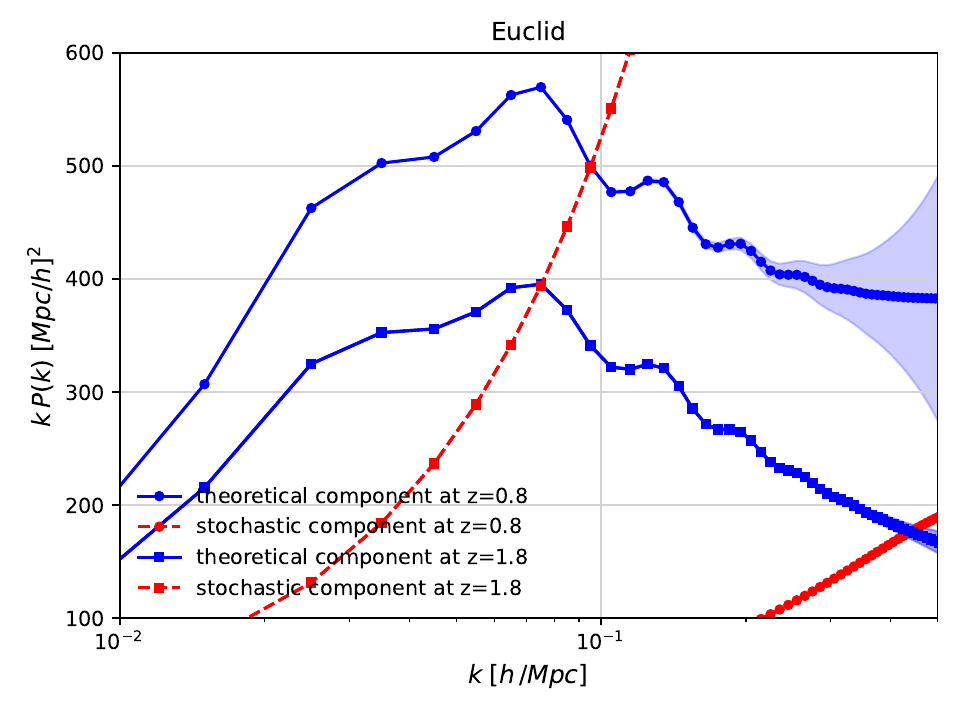}\includegraphics[scale=0.45]{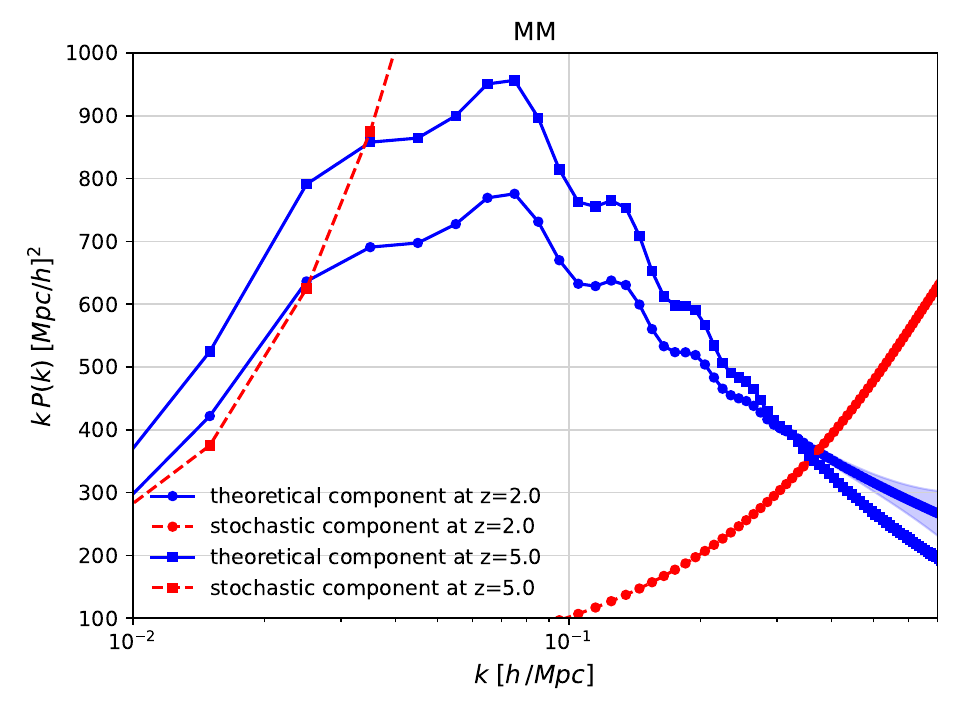}\\
\includegraphics[scale=0.45]{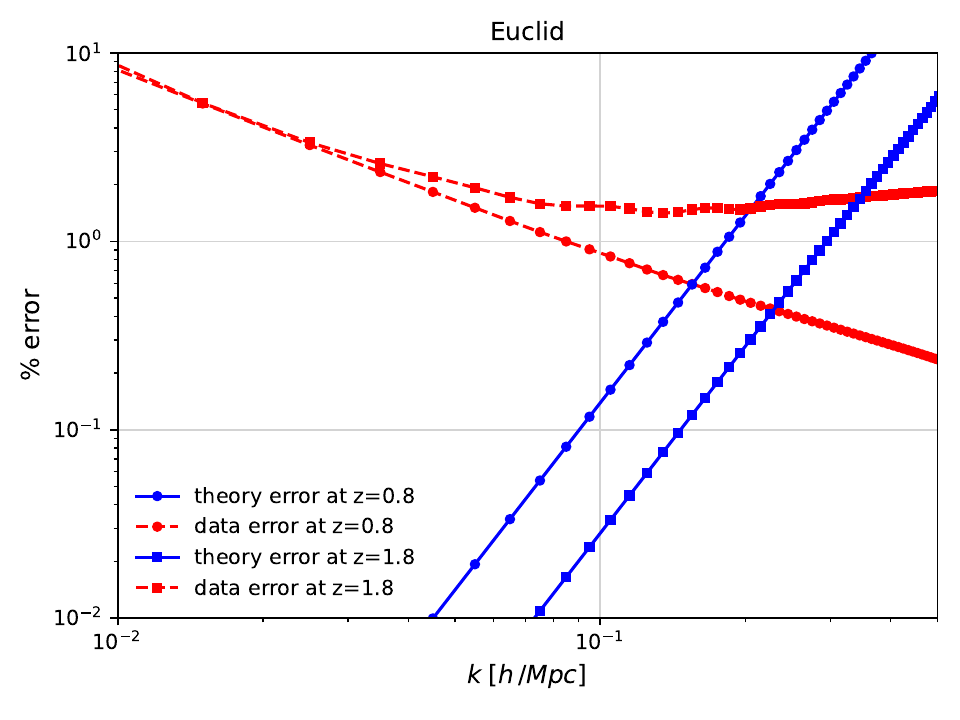}\includegraphics[scale=0.45]{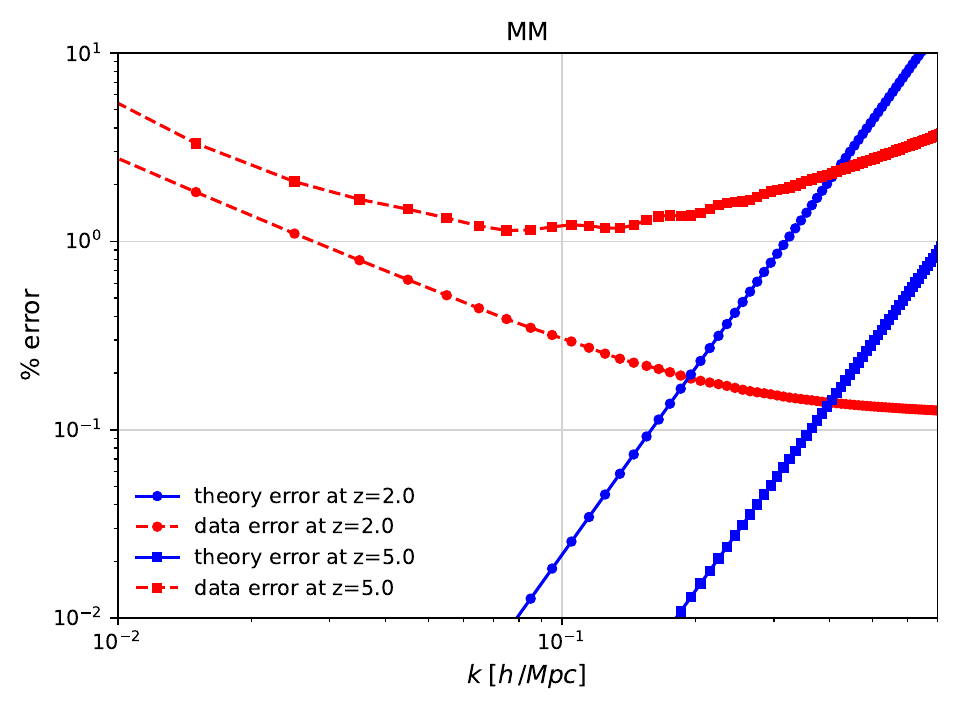}
\par\end{centering}
\caption{\label{fig:ps-noise}In the first row, we present plots showing the theoretical (solid, blue) and stochastic components (dashed, red) of the galaxy power spectrum for the Euclid and MM surveys at the first and last redshifts of their respective redshift ranges using fiducial bias parameters given in next subsection. For Euclid (figure on the left) we plot data at redshifts $z= 0.8$ (circle marker) and $z= 1.8$ (square marker). Similarly, for MM (figure on the right) we plot data at redshifts $z= 2.0$ (circle marker) and $z= 5.0$ (square marker). We also plot the theoretical error envelope as the `shaded' region over the theoretical curve. In the second row, we plot the expected data and theory errors for the aforementioned redshifts using Eqs.~(\ref{eq:CD}) and (\ref{eq:CE}) respectively. Note that the data error is dominated by cosmic variance (shot-noise) at low (high) k. The plots highlight that at the low redshifts, the signal-to-noise (S/N) is mainly constrained by the theoretical error rather than the stochastic component. At high redshifts, the signal is controlled by the stochastic component and bulk of the S/N is proportional to $b^2_1(z) n_g(z)$ (see \cite{Sailer:2021yzm} for a plot of $b^2_1(z) n_g(z)$ for the two surveys considered here.) }

\end{figure}

In Fig.~\ref{fig:ps-noise} we give plots of the theoretical galaxy power spectrum along with the expected stochastic component for the Euclid and MM surveys at the first and last redshifts for their respective redshift ranges aiming to highlight the factors affecting the expected signal. The signal-to-noise (S/N) ratio at each redshift primarily depends on two factors: the associated stochastic component, which is related to the number density of measured galaxies, and the theory error, which determines the reliability of theoretical estimates across different physical scales. When observing low redshifts with higher galaxy number densities, we anticipate that the S/N will be predominantly limited by the theoretical error. This observation is clearly evident in the plot shown on the left in the first row of Fig.~\ref{fig:ps-noise}. In the same figure, depicted in the second row, we present the expected data and theory errors by normalizing the square root of the data and theory covariances in Eqs.~(\ref{eq:CD}) and (\ref{eq:CE}) with the theoretical power spectrum. We find that the signal is mostly negligible for scales where the theory errors exceed a few $\%$. Consequently, we set a definitive cutoff, denoted as $k_{\rm{cut}}$, for the Fisher analysis, where the theory error reaches $\approx 15\%$.

In our forecast we do not explicitly model redshift measurement errors, however for both Euclid and MegaMapper these should not strongly affect our forecast, which relies on perturbative scales (see e.g. \cite{Sailer:2021yzm} for a discussion of redshift errors of these experiments). On the other hand, redshift space distortions (RSD) can affect our results and are not currently included in our forecast. We defer the inclusion of RSDs to future work.

\subsection{\label{subsec:Fiducial-parameter-values}Fiducial parameter values}

We set our baseline cosmological parameters to 
\begin{equation}
A  =1\hspace{24pt} \Omega_{b}  =0.0486 \hspace{24pt} \Omega_{c}  =0.2589 \hspace{24pt} n_{s}  =0.9667  \hspace{24pt} h  =0.6774
\end{equation}
where $A=A_{{\rm s}}/A_{{\rm s,fid}}$ for $A_{{\rm s,fid}}=2.1413\times10^{-9}$
and adopt uniform priors for these cosmological parameters from the
latest Planck analysis \citep{Planck:2018jri}.

Having set the fiducial values of the linear bias $b_{1}$ for each
survey as discussed in Sec.~\ref{subsec:Surveys}, we set the fiducial
values of remaining bias parameters as follows. For quadratic and
higher order tidal scalar biases, we consider a parametric form inspired
from Lagrangian bias or co-evolution models \citep{Chan_2012,Baldauf:2012hs,Eggemeier_2019}:
\begin{equation}
b_{2}=\frac{8}{21}\left(b_{1}-1\right),\qquad b_{\mathcal{G}_{2}}=-\frac{2}{7}\left(b_{1}-1\right),\qquad b_{\Gamma_{3}}=\frac{23}{42}\left(b_{1}-1\right).
\end{equation}
While the accuracy of the above functional form for setting up the
fiducial values is formally less than what we are working with in
the \eft, our final results (forecasts) are sufficiently insensitive
to these choices due to the maginalization over the bias parameters.
For instance, alternative fitting functions for quadratic bias $b_{2}$
obtained from N-body simulations and HOD modeling can be found in
\citep{Lazeyras:2015lgp,Yankelevich:2018uaz,DiDio:2018unb}.

Additionally, we will make the standard assumption that the scale-independent
stochastic contribution $b_{\epsilon}P_{\epsilon\epsilon}=P_{\rm{shot}}=1/\bar{n}_{g}$
and also set $B_{{\rm shot}}=1/\bar{n}_{g}^{2}.$ These control the
number of modes that will contribute to our isocurvature signal of
interest. If these numbers are larger, than the number of modes contributing
to the signal decreases. The fiducial values are aligned with standard
conservative practices, since detailed studies indicate the actual
values may be a bit smaller at large scales for massive halos, while
a positive correction can be expected for less massive halos \citep{Baldauf:2013hka}.

For the leading higher derivative and scale-dependent stochastic bias
parameters, $b_{\nabla^{2}\delta}$ and $b_{\nabla^{2}\epsilon}$, respectively,
we use the analysis of \citep{Eggemeier_2020} and assume following
fiducial values 
\begin{equation}
b_{\nabla^{2}\delta}=-1
\end{equation}
and 
\begin{equation}
b_{\nabla^{2}\epsilon}=-0.2
\end{equation}
with the clustering scale $k_{*}$ set as 
\begin{equation}
k_{*}\approx k_{{\rm HD}}\approx0.4\left(D_{+}(z)/D_{+}(0)\right)^{-4/3}\left({\rm h/Mpc}\right)\label{eq:kHD}
\end{equation}
where ${\rm HD}$ subscript stands for ``higher-derivative''. The particular choice of redshift dependence is motivated by a similar
expression for $k_{{\rm NL}}$ obtained from linear dimensionless
matter power spectrum. We note that the matter clustering scale $k_{*}$
need not have the same redshift dependence as the matter non-linear
scale $k_{{\rm NL}}$ \citep{Lazeyras:2019dcx}. However, we have
checked numerically that changing the redshift dependence does not
change the forecast results for the isocurvature signal, mostly because
we are marginalizing over the bias parameters. For simplicity, we
will present the results with the redshift dependence of Eq.~(\ref{eq:kHD}). We remark that the z-dependence of the fiducial value of the Laplacian bias differs from that of the non-Laplacian biases (such as $b_1$, $b_2$, $b_{\mathcal{G}_{2}}$, $b_{\Gamma_{3}}$, etc.), which have fiducial values inspired by the co-evolution models and thus scale as $\sqrt{1+z} \sim D_+(z)^{-0.5}$. Hence, unlike Laplacian biases, the fiducial values of the non-Laplacian biases increase with $z$. However, all non-Laplacian biases appear as prefactors of the one-loop terms given in Eq.~\ref{eq:PNLO} which have z-dependences given by the fourth power of growth function, $D_+(z)^4$. In contrast, the leading order Laplacian bias in Eq.~\ref{eq:P_lap} multiples the linear power spectrum which scales as $D_+(z)^2$. Consequently, the one-loop terms contain an additional quadratic power of the growth function. Thus, despite of the non-Laplacian biases increasing with $z$, the effective z-scaling (including biases) of the one-loop terms in Eq.~\ref{eq:PNLO} and the leading order Laplacian contribution in Eq.~\ref{eq:P_lap} is almost the same. Both of these terms vanish (become insignificant) on large scales at a similar rate at higher redshifts.

Finally, for the purposes of forecasting we will consider the following
reduced set of nuisance parameters:
\begin{equation}
\{\mbox{nuisance}\}=\{b_{1},b_{2},b_{\mathcal{G}_{2}},b_{\nabla^{2}\delta},b_{\nabla^{2}\epsilon}\}\label{eq:Pset}
\end{equation}
where we note that since the pair of bias parameters $\{b_{\mathcal{G}_{2}},b_{\Gamma_{3}}\}$
and $\{b_{1}c^{2},b_{\nabla^{2}\delta}\}$ are known to be highly
degenerate, we do not include $b_{\Gamma_{3}}$ and $b_{1}c^{2}$
within our set of nuisance parameters over which we vary the likelihood
function. The reason for the degeneracy of $\{b_{1}c^{2},b_{\nabla^{2}\delta}\}$
is small $k$ expansion, while the reason for the degeneracy of $\{b_{\mathcal{G}_{2}},b_{\Gamma_{3}}\}$
seems to be an accidental symmetry of the convolution existing at
this order in perturbation theory. Hence, the Fisher matrix information
does not include theoretical derivatives with respect to these two
model parameters. We set Gaussian priors for the bias parameters with
the standard deviations 
\begin{equation}
\sigma_{b_{1}}=4,\quad\sigma_{b_{2}}=2,\quad\sigma_{b_{i}}=1
\end{equation}
which apply to bias parameters in Eq.~(\ref{eq:Pset}). Note that
we do not include the prior for a bias when we marginalize over it.

For the Fisher analysis (see Appendix \ref{sec:Fisher-matrix-and}
for a review of Fisher forecast formalism), we set a coarse momentum
bin width $k_{{\rm bin}}=0.01\,{\rm hMpc^{-1}}$ with $k_{\min}=0.005\,{\rm hMpc^{-1}}$
and error envelope correlation length $\Delta k=0.1\,{\rm hMpc^{-1}}$.
We found that the results change by less than 10\% to the choice of
$\Delta k$ in the range $0.05-0.13\,{\rm hMpc^{-1}}$. We set the
pivot scale at $k_{{\rm p}}=0.05\,{\rm Mpc^{-1}}$, following the Planck convention, such that the results
for the ratio of the amplitude of the primordial isocurvature to adiabatic
power spectrum $\alpha$ are given at $0.05\,{\rm Mpc^{-1}}$.

\section{\label{sec:Results-and-Discussion}Results and Discussion}

\begin{figure}
\begin{centering}
\includegraphics[scale=0.7]{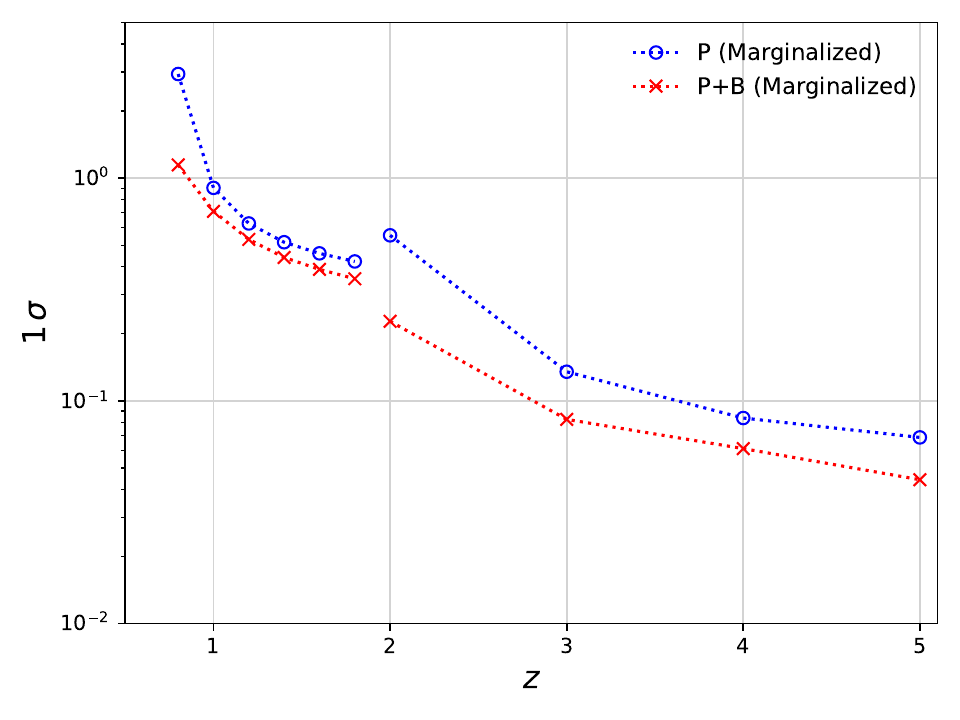}
\par\end{centering}
\caption{\label{fig:1--error-forecast}1-$\sigma$ error forecast for an example
isocurvature amplitude $\alpha=0.01$ and spectral index $n_{{\rm iso}}=3$,
achievable by Euclid and MegaMapper (MM) as a function of the redshift
reach. Here, the curves for $z\lesssim2$ correspond to those for
Euclid while the curves for $z\gtrsim2$ correspond to those for MM,
and for example the point at $z=$5 accounts for the information accumulated
in the redshift range 2 to 5. The decrease in the error results from
a combination of both the decreasing theoretical uncertainty as well
as the increase in the number of modes available at higher redshifts
(i.e.~number of modes increase with $k_{\mathrm{max}}^{3}$ where
$k_{\mathrm{max}}$ represents the maximum $k$-mode with sufficiently
accurate theoretical predictions as well as with the number of redshift
bins). The label $P$ denotes that the one-loop galaxy power spectrum
and the label $B$ denotes the tree-level bispectrum are used in making
the forecasts. Note that most of the constraining power comes from
the power spectrum because of the limited number of modes available
for the tree-level bispectrum compared to a one-loop power spectrum.
The inclusion of the bispectrum signal helps to break the degeneracy
(correlation) from $b_{1},b_{2},$ and $b_{\mathcal{G}_{2}}$ bias
parameters and thus reduces the error by $\approx20-30\%$ compared
to a power spectrum only analysis.}
\end{figure}

Fig.~\ref{fig:1--error-forecast} shows marginalized\footnote{marginalized over all bias/nuisance parameters}
1-$\sigma$ error sensitivity for the isocurvature amplitude, $\alpha$
, achievable by Euclid and MegaMapper experiments as a function of
the redshift reach. Larger the redshift, more modes are available
for the one-loop power spectrum since the error envelope is smaller
at higher redshifts. The resulting cumulative signal sensitivity at
higher redshifts is responsible for the decreasing error shown in
the plot. We expect the error forecast to be smaller in magnitude
in the idealized galaxy bias scenario as discussed below Table \ref{tab:MM}.
Hence, this forecast should be seen as a conservative estimate, and
we would realistically expect better results. Furthermore, since there
is no RSD in the observables considered here, its inclusion can break
the bias degeneracies and improve signal in a fuller analysis. (See
\citep{Chudaykin:2019ock} for a recent forecast on neutrino masses
that highlights the improvement in error bars from RSD and AP effects.)

Most of the constraint/signal in Fig.~\ref{fig:1--error-forecast}
comes from the one-loop power spectrum due to the larger number of
modes available compared to the tree-level bispectrum. However, the
inclusion of the bispectrum signal helps to break the degeneracy (correlation)
from $b_{1},b_{2},$ and $b_{\mathcal{G}_{2}}$ bias parameters to
some extent and thus reduce the error by $\approx20-30\%$ compared
to a power spectrum only analysis. This is illustrated by `$\times$'
marked (red) curve. If the bispectrum computation is improved, then
one would naively expect further significant improvements. However,
because of the proliferation of the bias parameters at one-loop bispectrum
level, there is a trade-off whose final effects to the error computation
is not obvious \citep{Eggemeier_2019,Philcox:2022frc}. There is some
previous work \citep{Eggemeier_2021} which argues that these higher
order bias contributions to the bispectrum can be fixed using co-evolution
and peak background split relations (\citep{Lazeyras:2015lgp}) such
that the required number of independent bias parameters is greatly
reduced and thus do not completely degrade the gain from the improved
accuracy. 
\begin{figure}
\centering{}\includegraphics[scale=0.7]{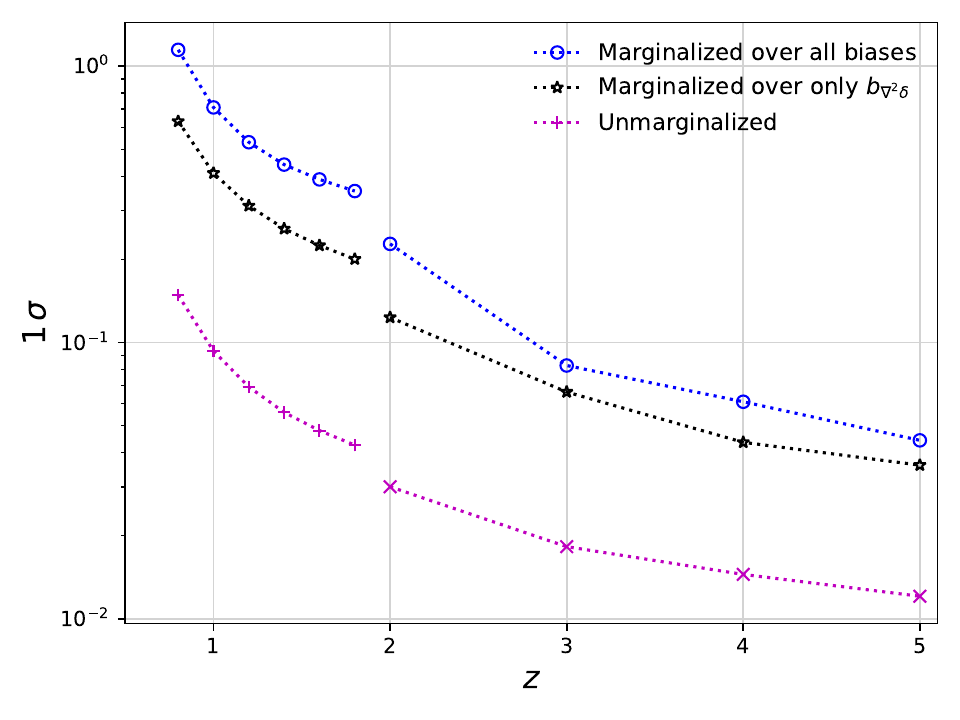}\caption{\label{fig:marginalization}This figure is similar to that of Fig.~\ref{fig:1--error-forecast}
except only the combined power spectrum and bispectrum signal has
been used and more importantly, the bias parameter marginalization
has been turned on and off to understand the sensitivity of the signal
to the bias parameters. As can be seen in the star marked curve, most
of the marginalization induced signal degradation at high $z$ is
coming from the marginalization over the Laplacian bias parameter
$b_{\nabla^{2}\delta}$.}
\end{figure}

In Fig.~\ref{fig:marginalization}, we illustrate the effects of
marginalization over the bias parameters on the $\alpha$. There is
about a factor of 10 (5) change in the sensitivity arising from the
combination of five bias parameters ($b_{1},b_{2},b_{\mathcal{G}_{2}},b_{\nabla^{2}\delta},b_{\nabla^{2}\epsilon}$)
that can affect the galaxy count similarly as a primordial isocurvature
power spectral component in the case of Euclid (MM). Note that $b_{\nabla^{2}\delta}$
and $b_{\nabla^{2}\epsilon}$ parameters are multiplying higher power
dependence in $k$ than others and thus lead to an approximate degeneracy
of the high spectral power component coming from the isocurvature
perturbations. Additionally due to the lack of $b_{\nabla^{2}\delta}$
and $b_{\nabla^{2}\epsilon}$ bias parameters in the tree-level bispectrum
signal, the degeneracy from these biases, especially $b_{\nabla^{2}\delta}$,
can become stronger than others. The degeneracy from $b_{\nabla^{2}\delta}$
is stronger than the degeneracy from $b_{\nabla^{2}\epsilon}$ because
of its closer similarity of the associated $k$-dependent term with
the isocurvature contribution for $n_{\rm{iso}}\sim3.5$ (see Eq.~(\ref{eq:P_lap}) in contrast
with Eq.~(\ref{eq:ggeps}) where $P_{\epsilon\epsilon}$ is a constant)\footnote{The leading power spectrum contribution from the isocurvature amplitude
is given as
\begin{align}
P_{gg}(k) & \supset b_{1}^{2}P_{11}^{{\rm AD}}(k)\alpha\left(\frac{f_{c}}{3}\right)^{2}\left(\frac{T_{{\rm iso}}(k)}{T_{{\rm ad}}(k)}\right)^{2}\left(\frac{k}{k_{\mathrm{p}}}\right)^{n_{{\rm iso}}-n_{{\rm ad}}}\\
 & \propto b_{1}^{2}\alpha P_{11}^{{\rm AD}}(k)\left(\frac{k}{k_{\mathrm{p}}}\right)^{n_{{\rm iso}}-n_{{\rm ad}}-0.5}
\end{align}
where $\lim_{k>0.1{\rm h/Mpc}}\left(T_{{\rm iso}}(k)/T_{{\rm ad}}(k)\right)^{2}\propto k^{-0.5}$. The above spectral shape is similar
to the leading contribution from the Laplacian bias $b_{\nabla^{2}\delta}$:
\begin{equation}
P_{gg}(k)\supset-b_{1}b_{\nabla^{2}\delta}P_{11}(k)k^{2}
\end{equation} 
for values of $n_{\rm{iso}}$ close to $3.5$. The total linear power spectrum in the above expression is given in Eq.~(\ref{eq:Pm_analytical}).}. This is highlighted in the star marked curve in Fig.~\ref{fig:marginalization}
where we show the effect of marginalizing over only the Laplacian bias
$b_{\nabla^{2}\delta}$. Higher loop computations as well as additional
observables such as RSD will help reduce the degradation from the
marginalization. The fact that Euclid shows a greater degradation
from marginalization can be attributed to its lower redshift range
which limits the $k$ range over which the signal is obtained due
to the error envelope (e.g.~a fixed $k$ value signal can be mimicked
by almost any bias parametric degeneracy while a large $k$ range
signal will not be reproducible exactly by the bias dependent spectral
function). The decrease in degradation from the marginalization when
accounting for a larger redshift range (see the marginalization
over $b_{\nabla^{2}\delta}$ in Fig.~\ref{fig:marginalization})
may be attributable to the redshift dependence of the biases, which is more constraining when probed over a larger range of redshifts.
%whose normalization can always be adjusted at one redshift bin and would then be constraining for higher redshift bins.

\begin{figure}
\begin{centering}
\includegraphics[scale=0.5]{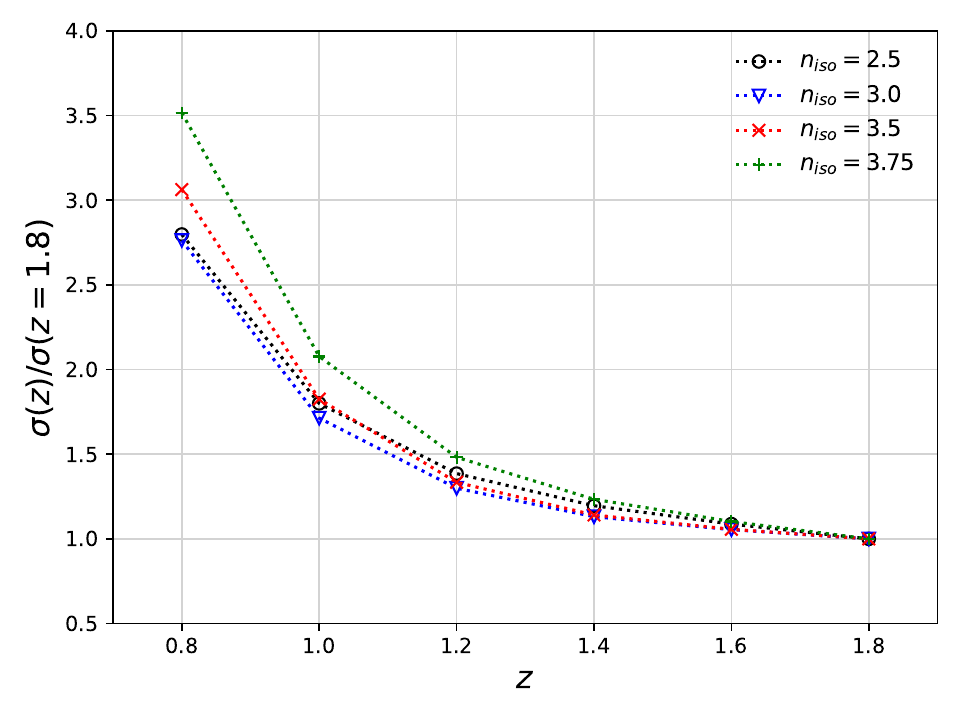}~\includegraphics[scale=0.5]{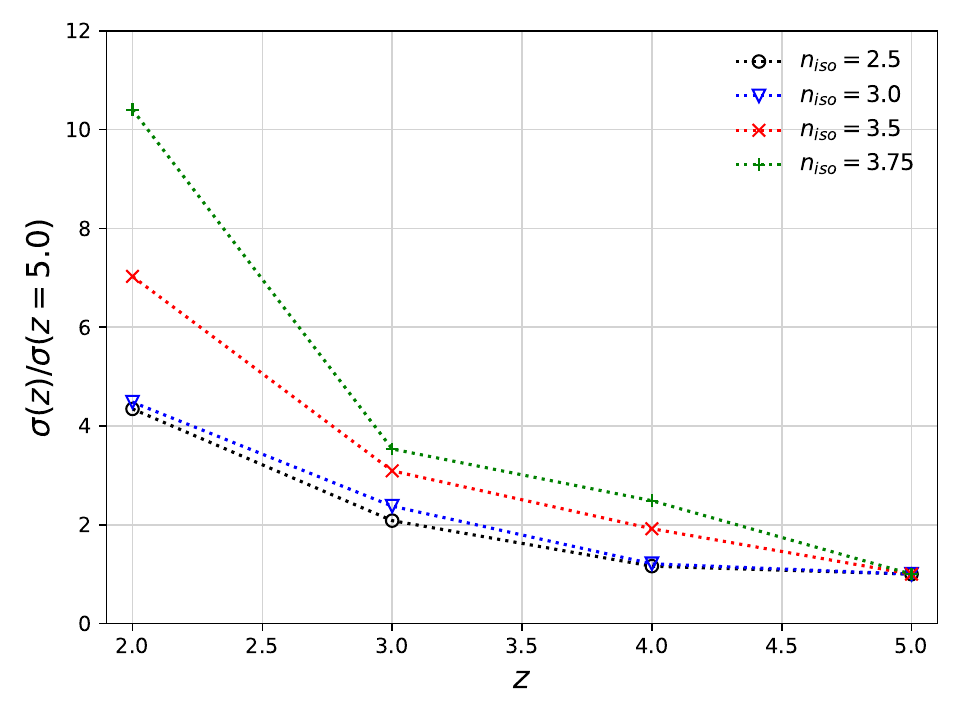}
\par\end{centering}
\caption{\label{fig:spectraldep}The dependence of the errors (Euclid on the
left and MM on the right) on the isocurvature spectral index $n_{\mathrm{iso}}$
are shown. Only the power spectrum is used to make these plots since
it is the dominant contribution to the sensitivities. Because MM is
able to probe a larger number of high $k$ modes (due to the high
redshift yielding more accurate theoretical computation), the spectral
dependence is stronger for MM.}
\end{figure}

Fig.~\ref{fig:spectraldep} illustrates the isocurvature spectral
dependence of the sensitivities of Euclid and MM. Because MM observes
a higher redshift universe where the perturbations have not grown
as much, it has less theoretical error on large $k$ scales. This
means that the degeneracies coming from the bias with high $k$ spectral
dependence can vary more dramatically in the case of MM compared to
Euclid, as seen in the figure. The decrease in the error with increasing
$z$ seems to be continuing in the case of $n_{\mathrm{iso}}=3.75$
with MM, while for smaller spectral indices, there is an indication
of saturation.

\begin{figure}
\begin{centering}
\includegraphics[scale=0.8]{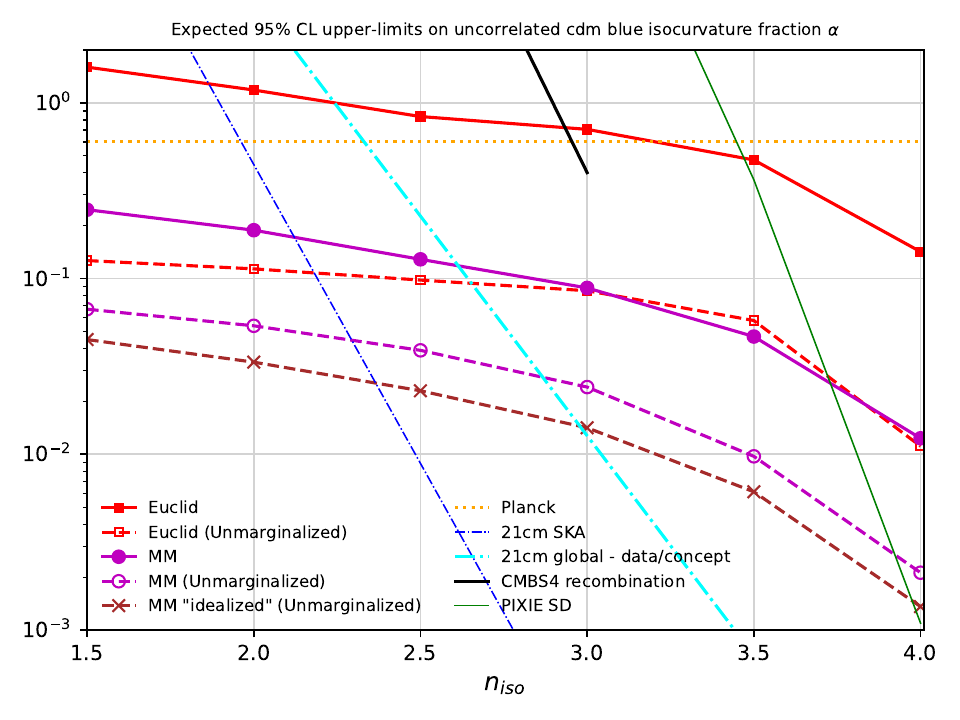}
\par\end{centering}
\caption{\label{fig:moneyplot}Our forecast for the Euclid and MM 2-sigma upper
limits that can be placed on the isocurvature power spectral amplitude
fraction $\alpha$ is shown as a function of the isocurvature spectral
index. The results from Euclid and MM are drawn using \textquotedblleft square\textquotedblright{}
and \textquotedblleft circular\textquotedblright{} markers respectively
where solid (open) markers reflect marginalized (unmarginalized) errors.
Because we are not including effects such as RSD or including higher
loop computations, the actual upper limit should be somewhere between
the unmarginalized and marginalized curves. For comparison, we also
plot (without markers) some of the upper limits existing in a sample
of the literature. The dotted (orange) line represents the current
2-sigma upper bound from Planck \citep{Planck:2018jri}. The solid
thin (green) and thick (black) curves are obtained from \citep{Chluba_2013}
and \citep{Lee:2021bmn} respectively. From previous forecasts based
on 21cm line analyses, we plot thin (blue) and thick (cyan) dashdot
curves obtained from \citep{Sekiguchi:2013lma,Takeuchi:2013hza} and
\citep{Minoda:2021vyw} respectively.}
\end{figure}

In Fig.~\ref{fig:moneyplot}, we forecast the 95\% confidence upper
bound constraint on the uncorrelated CDM isocurvature fraction $\alpha$
for Euclid and MM. The fully marginalized Euclid errors for $\alpha$
are below the current Planck upper bound for $n_{\mathrm{iso}}\gtrsim3$,
and the fully marginalized MM errors are below the current Planck
upper bound for $n_{\mathrm{iso}}\in[1.5,4]$. This result can be
interpreted as the expected improvement achievable in the near future
experiments. Because we have conservatively taken the linear bias
from the ``fiducial'' categorization seen in Table 2 of \citep{Ferraro:2019uce},
the actual experimental sensitivity may be somewhat better than what
we see in the solid curves with markers. Note that marginalized MM
case is as good as unmarginalized Euclid case for the reasons described
above.

Because we have not broken some parametric degeneracies using all
available observables such as the RSD, to gain intuition on how much
smaller the errors become with the degeneracy breaking, we also plot
in Fig.~\ref{fig:moneyplot} the unmarginalized\footnote{Here, we are mainly interested in not marginalizing over the biases.}
case shown in dashed curves with markers. To understand the limitations
imposed by the ``fiducial'' bias, we also plot (`$\times$' marked)
the unmarginalized forecast for MM with the idealized linear biases
and galaxy densities obtained from table 1 of \citep{Ferraro:2019uce}.

To give a sense of how the expected improvements compare with other
future constraints, we sample results from the literature:
\begin{itemize}
\item In \cite{Planck:2018jri} the authors analyzed the Planck data to constrain the isocurvature amplitude for various CDI models. For the uncorrelated ADI+CDI with free $n_{\rm{iso}}$ (axion-II) model they give the upper bound of $\alpha \lesssim 0.6$ at $k_{\rm{p}}=0.05$ Mpc$^{-1}$ over a range of spectral indices $1.55<n_{\rm{iso}}<3.67$. Since the Planck data is restricted to scales $k\lesssim 0.1$ Mpc$^{-1}$ where the isocurvature power hasn't grown much, the data was not very constraining on the isocurvature spectral index and hence we consider a flat constraint over the range of indices considered in this work.
\item Compared to the CMBS4 recombination forecast of \citep{Lee:2021bmn}
and PIXIE spectral distortion forecast \citep{Chluba_2013}, our results
are obtained from much smaller $k$ values. For example, because \citep{Lee:2021bmn}
gave their results from 30 Mpc$^{-1}$, we extrapolated their results
down to our $k$ range (with a pivot scale of $0.05$Mpc$^{-1}$)
for comparison. There is a break in the black solid curve at $n_{\mathrm{iso}}=3$
because \citep{Lee:2021bmn} has not given their results beyond that
spectral index.
\item \citep{Sekiguchi:2013lma} gives a forecast on $\alpha$ from SKA
based on 21cm emission from minihalos in the $z\in[5,20]$. Because
the halo mass function used utilizes small virialized objects, these
constraints are sensitive to nonlinear physics at $k\sim O(100)$Mpc$^{-1}$.
Because $\alpha$ of \citep{Sekiguchi:2013lma} is defined at the
pivot scale of $k=0.002$Mpc$^{-1}$ and because their definition
is the total isocurvature fraction while our $\alpha$ definition
is the CDM isocurvature fraction, we have scaled their results appropriately
in Fig.~\ref{fig:moneyplot}. The results of \citep{Takeuchi:2013hza}
are also similar. Note that if one uses the strongly nonlinear observables
such as the number counts of virialized objects, then $n_{\mathrm{iso}}$
can be strongly constrained \citep{Takeuchi:2013hza}.
\item In \citep{Minoda:2021vyw}, constraints are placed on the isocurvature
fraction $\alpha$ coming from the shift in the 21cm absorption trough
redshift which increases as $\alpha$ increases. In our plot, we give
the bound coming from the assumption that Edges signal sees a trough
starting at $z_{\mathrm{min}}=17.2$. Note that their equation 2.3
is inconsistent with its usage in equation 2.6 since with their 2.6,
their equation 2.3 should have an extra factor of the square of CDM
fraction and a factor of $1/3$ coming from the map of the primordial
isocurvature perturbation to the matter power spectrum as discussed
in Sec.~\ref{sec:Mixed-power-spectrum}.\footnote{Recently, an interesting paper \cite{Esteban:2023xpk} appeared which gives similar constraints to the "21cm global - data/concept" curve in Fig.~\ref{fig:moneyplot}.  The analysis in that paper is based on milky way satellite galaxies and their results are sensitive to the nonlinear physics at comoving scales in the range of 5 Mpc$^{-1} < k < $  40 Mpc$^{-1}$, which means a more realistic isocurvature spectrum that has a cutoff before 5 Mpc$^{-1}$ will be able to evade that paper's constraints while the constraints of our Fig.~\ref{fig:moneyplot} will still apply.}
\end{itemize}
It is clear that near future experiments will realistically give at
least an order of magnitude improvement in the constraint of the isocurvature
amplitude for high spectral indices, even with the current theoretical
computational technology of structure formation.

\section{\label{sec:Conclusion}Conclusion}

We have presented a Fisher forecast of the Euclid and MegaMapper experimental
sensitivities to primordial blue isocurvature spectrum of the uncorrelated
CDM type. Our analysis utilized \eft (which uses SPT as part of its
machinery) for the density field evolution because of its better UV
control through N-body matched renormalization prescription and perturbative
control of the uncertainties. This was then used with the bias expansion
to obtain the theoretical prediction for galaxy number counts. We
find that Euclid can give a factor of few improvement on the isocurvature
spectral amplitude compared to the existing Planck constraints for
$n_{\mathrm{iso}}\gtrsim3$ while MM can give about 1 to 1.5 order
of magnitude improvement for a broad range of $n_{\mathrm{iso}}$
(Fig.~\ref{fig:moneyplot}). MM tends to give a better constraint
because of the larger signal data volume. Also, as expected, most
of the signal is coming from the power spectrum and not the bispectrum
for these experiments as can be seen in Fig.~\ref{fig:1--error-forecast}.
Our forecast here represents conservative estimates, and an optimistic
characterization (an $m_{UV}=24.5$ magnitude-limited dropout sample) of the experiments can improve the situation by an
order of magnitude.

Going to higher redshift changes the isocurvature amplitude measurement
sensitivities more significantly for higher spectral indices because
of the larger amplitude that exists at larger $k$ values and higher
redshift results in smaller error envelope for any given $k$ value.
When comparing Euclid and MM, this effect is more significant for
MM as can be seen in Fig.~\ref{fig:spectraldep} because
MM has a higher redshift reach. As can be seen in Fig.~\ref{fig:marginalization},
the degradation in the constraint due to the bias marginalization
is dominated by $b_{\nabla^{2}\delta}$ coming from the spectral similarities
of this contribution when compared with an isocurvature spectrum of
$n_{\mathrm{iso}}\sim3.5$. This degeneracy does not seem to exist in
scenarios without the large blue isocurvature contribution \citep{DAmico:2019fhj}.

One interesting theoretical feature of the isocurvature scenario in
the \eft formalism is that the bare sound speed parameter $c^{2}$
is both cutoff dependent and may even become negative when the cutoff
is taken above $\Lambda\gtrsim O(3)$ Mpc$^{-1}$ for sizable isocurvature
amplitudes. Nonetheless, since physical observables are dependent only
on the renormalized parameter $c_{{\rm ren}}^{2}$, we compare renormalized
parameter $c_{{\rm ren}}^{2}$ for a sample of primordial spectra
in Fig.~\ref{fig:delta_c2ren}. We see in the figure that the difference
in the renormalized value of the effective sound speed squared $c_{{\rm ren}}^{2}$
between mixed and pure ICs at redshifts $z=0$ and $1$ obtained from
the N-body simulations fits well a semi-analytic fitting formula
for $\Delta c_{{\rm ren}}^{2}$ given in Appendix \ref{sec:Semi-analytic-expression-for-csrenMX}
which is applicable for values of $O(0.01)\lesssim\alpha<1$ and $2\leq n_{{\rm iso}}\leq4$
for redshifts $z\leq2$. Since the fluid parameter $c^{2}$ is degenerate
with the higher order derivative bias $b_{\nabla^{2}\delta}$, a precise
value of $c_{{\rm ren}}^{2}$ for mixed ICs is not required as we
marginalize over the bias parameters. Hence, we found it adequate
to work with the renormalized EFT or bias parameters of adiabatic
ICs and apply them directly to the analysis/forecasts for mixed IC
cosmologies.

Another peculiarity of the large blue spectral isocurvature scenario
is the renormalization of various one-loop contributions to the galaxy
power spectrum as discussed in Sec.~\ref{sec:UV-div-and-corr} and
the individual bare contributions given in Appendix \ref{sec:Galaxy-one-loop-contributions}.
Unlike the case for pure adiabatic ICs, the UV sensitivity for mixed
ICs with large blue spectral indices is more severe since the one-loop
terms $\mathcal{I}_{\delta^{(2)}\delta^{2}}(k)$, $\mathcal{I}_{\delta^{2}\delta^{2}}(k)$
and $\mathcal{I}_{\delta^{2}\mathcal{G}_{2}}(k)$ are UV divergent
when $n_{{\rm iso}}\gtrsim3.5$ and $\mathcal{F}_{\mathcal{G}_{2}}(k)$
is divergent when $n_{{\rm iso}}\gtrsim3$. The set of terms $\mathcal{I}_{\delta^{(2)}\delta^{2}}(k)$,
$\mathcal{I}_{\delta^{2}\delta^{2}}(k)$ and $\mathcal{I}_{\delta^{2}\mathcal{G}_{2}}(k)$
are renormalized by the higher order derivatives of stochastic terms
in the bias expansion, and similarly the leading divergent piece from
$\mathcal{F}_{\mathcal{G}_{2}}(k)$ is renormalized by redefining
the $b_{\nabla^{2}\delta}$ bias parameter of the leading higher-derivative
operator. The renormalized one-loop contributions are given in Eq.~set (\ref{eq:renormalized_PNLO}).

Fisher forecasts have well known limitations. For example, the assumption
of a Gaussian likelihood for the Fisher analysis may break down at
scales comparable or much smaller than nonlinear scales due to loop
corrections as well as one-halo and higher order perturbative terms.
Similarly, we have also neglected the small correlations among the
redshift bins for each of the surveys. Additionally the
parameter degeneracies in a Fisher analyses can be misleading since
they are only considered at leading order by construction \citep{Ryan:2022qpa}.
Beyond the limitations coming from the Fisher formalism, there are
limitations in the bias functions used here. In this work,
we have used the fiducial values for higher order Laplacian biases
\textbf{$b_{\nabla^{2}\delta}$ }and \textbf{$b_{\nabla^{2}\epsilon}$}
from the literature wherein the ICs were taken to be purely adiabatic
because for small isocurvature amplitudes $\alpha\sim O(0.01)$, we
do not expect the renormalized biases to be very different from those
of the adiabatic case, and marginalization seems to remove dependences
upon the exact fiducial values of the biases. In summary we do not expect any of these limitations to strongly affect our forecast.
%Nonetheless, a strong prior and a more accurate fiducial set should result in better forecast results.

There are many interesting future directions related to this work. The most immediate extension of our work will be to include RSDs. RSD effects can break some of the bias degeneracies significantly
since these include velocity fields such that each multipole provides
an additional observable, wherein the biases have varying sensitivities
to these multipoles. For example, \cite{Chudaykin:2019ock} showed that RSD significantly improves constraints on neutrino masses, and we expect that a similar improvement is possible in the case of blue isocurvature perturbations.

% There are other many future directions related to this work. Perhaps
% the strongest improvement in the error forecast would come from including
% more observables to break the parametric degeneracies with the bias. 
% In particular, RSD effects can break some of the bias degeneracies significantly
% since these include velocity fields such that each multipole provides
% an additional observable wherein the biases have varying sensitivities
% to these multipoles \moritz{add cite}. It is encouraging that these future experiments
% seem to be sensitive to the blue isocurvature scenarios even with
% the current conservative analysis.

One important theoretical caveat to the scenario that we are considering is that the primordial isocurvature spectrum is assumed to rise
indefinitely without a cutoff. In reality, one can show \citep{Chung:2015tha}
that when the spectral index is larger than about 2.4, the energy
density in the dark matter gets diluted away by the inflationary expansion.
Hence, a more realistic spectrum should have a cutoff and become flat
as in the model of \citep{Kasuya:2009up}. This can lead to a break
in the degeneracy with the bias parameters. Furthermore, as shown
in \citep{Chung:2021lfg}, when the underlying dynamics that generates
the blue spectrum is underdamped, a large (an amplitude multiplicative
enhancement of $O(30)$) oscillatory features can arise in the primordial
isocurvature power spectrum. Such an oscillatory spectrum will also
tend to break the parametric degeneracies differently than the case
of the power law primordial spectrum considered in the present work.
A future work dedicated to the study of these scenarios would be useful.

It would be interesting to include massive neutrinos in the theoretical
prediction to see how much the neutrino mass constraints degrade due
to the presence of blue CDM isocurvature component which can qualitatively
compensate for the power erasure due to the neutrino free streaming.
Although naively we do not expect a quantitatively precise degeneracy
between $\alpha$ and neutrino masses since neutrino suppression is
roughly constant at small scales unlike scale-dependent blue isocurvature
signals, a low-$n_{{\rm iso}}$ isocurvature signal can cancel some
suppression from a massive neutrino at quasi non-linear scales.

Another natural target of exploration is to go beyond one-loop power
spectrum and tree-level bispectrum to explore possible trade-offs
between larger number of available modes and proliferation of bias
parameters. Previous analyses from the literature \citep{Eggemeier_2019,Philcox:2022frc}
suggest that the expected improvement is paltry due to the marginalization
over an extended set of biases unless strong priors can be placed
on the bias parameters. Stronger priors on the values and covariance of bias parameters could in principle be obtained from simulations, if these simulations can bracket the range of physically plausible small-scale physics. A different strategy to break parameter degeneracies is to perform a field level analysis rather than to calculate N-point correlation functions. That field level analysis can break bias parameter degeneracies was recently shown in \cite{Baumann:2021ykm} for the case of primordial non-Gaussianities. Yet another possible alternative is to come up with a non-perturbative
description of tracers in the case of mixed initial condition/high
spectral index situations which is unlikely to have the same degeneracy
as the perturbative expansion. For example, blue isocurvature can strongly affect the halo mass function. We hope to explore some of these directions in the future.

In addition, we observe from Fig.~{\ref{fig:ps-noise}} that for the low redshift surveys with large number densities, the signal is primarily limited by the theoretical error. In Fig.~{\ref{fig:eft-fastpm-compare}} we noted that the two-loop error envelope in Eq.~(\ref{eq:error}) is somewhat conservative and seems to be overestimating the actual error between the one-loop EFT and the N-body data.  Consequently, obtaining a more accurate estimation of the theory error specifically for the biased tracers could potentially result in improved forecast and fitting outcomes in future analyses.

Finally, since our Eq.~(\ref{eq:dcs2_final}) was derived for $z\lesssim2$,
it would useful to more accurately compute $\Delta c_{\mathrm{ren}}^{2}$
for higher redshifts (for example in the range $2<z<10$) if there
are more observables to break its degeneracy with the biases. The
current fitting function from N-body data suggests that the ratio
of $\Delta c_{{\rm ren}}^{2}/c_{{\rm ren,AD}}^{2}\propto D_{+}^{1.2}(z)$
increases almost linearly with redshift. This can be explained if
we consider that the renormalized parameter $c_{{\rm ren}}^{2}$ inherits
UV effects due to the non-linear clustering at small scales. The non-linear
scale $k_{{\rm NL}}$ increases with redshift $z$, and hence the
difference in the linear power spectra $P_{11}^{{\rm MX}}(k)-P_{11}^{{\rm AD}}(k)$
at $k=k_{{\rm NL}}$ also increases for $n_{{\rm iso}}>n_{{\rm ad}}$.
Consequently, we expect a larger power from mixed ICs at small scales
which may explain why the ratio $\Delta c_{{\rm ren}}^{2}/c_{{\rm ren,AD}}^{2}$
increases with redshift $z$. On the other hand, given that the bare
$c^{2}$ can become negative for $\Lambda\gtrsim O(3)$, there is
an unclear interpretation of this parameter, which leaves room for
more intricate UV dynamics being at play here for the isocurvature
scenario. A nonlinear UV model exploration of this issue may be useful
to further elucidate the differences in $c_{{\rm ren,MX}}^{2}$ and
$c_{{\rm ren,AD}}^{2}$.
\begin{acknowledgments}
We thank Utkarsh Giri for teaching us how to use the numerical tools
and helpful discussions. TSC thanks Adrian Bayer for discussions on
nbodykit. TSC was supported in part by Wisconsin Alumni Research Fund
at the University of Wisconsin-Madison.  DJHC acknowledges partial support from DOE grant DE-SC0017647.  MM acknowledges support from DOE grant DE-SC0022342.
\end{acknowledgments}

\appendix

\section{\label{sec:Renormalization-scheme}Renormalization scheme}

In a perturbatively expanded effective field theory valid in the IR
and coarse grained over a UV inverse length scale $\Lambda$, one
writes down all possible terms consistent with the symmetries up to
a given perturbative order of computation and adjusts the coefficients
of the terms using  a chosen prescription called a renormalization
scheme. In this section, we list the renormalization scheme used for
the \eft and the bias expansion for this paper.

Let there be a set of (composite) operators $\{\mathcal{O}_{i}\}$
that can mix because they share the symmetry representation of the
theory. Note that $\mathcal{O}_{i}$ can involve gravitational potentials
which in the density fields can be nonlocal. In the EFTofLSS, some
of the $\mathcal{O}_{i}$ are defined using expressions obtained from
a derivative expansion. Define the renormalized operator $\left[\mathcal{O}_{i}\right]$
as \citep{Assassi:2014fva}\footnote{Other schemes such as multi-point propagator methods have been recently
discussed in light of renormalizing the bias operators and coefficients
\citep{Eggemeier_2019}} 
\begin{equation}
\left[\mathcal{O}_{i}\right]=\mathcal{O}_{i}+\sum_{j}Z_{ij}(\Lambda)\mathcal{O}_{j},\label{eq:CO_expn}
\end{equation}
where $Z_{ij}$ contain the counterterm coefficients appropriately
chosen to make the correlators involving $\left[\mathcal{O}_{i}\right]$
independent of $\Lambda$. To determine $Z_{ij}(\Lambda)$, \citep{Assassi:2014fva}
proposes a natural sufficient condition that all $n$-point correlation
functions be finite in the large-scale limit $(k\rightarrow0)$. This
is equivalent to the following renormalization prescription for the
composite operators $\mathcal{O}_{i}$:
\begin{equation}
\left\langle [O_{i}](\vec{k})\delta^{(1)}(\vec{k}_{1})\cdots\delta^{(1)}(\vec{k}_{n})\right\rangle =\left\langle O_{i}(\vec{k})\delta^{(1)}(\vec{k}_{1})\cdots\delta^{(1)}(\vec{k}_{n})\right\rangle _{{\rm tree}}\qquad\forall k_{i}\rightarrow0\label{eq:renom_cond}
\end{equation}
where $\delta^{(1)}(k)$ is the linear matter over-density. This implies
for Eq.~(\ref{eq:CO_expn}) the condition that the loop contributions
cancel since
\begin{equation}
\left\langle O_{i}(\vec{k})\delta^{(1)}(\vec{k}_{1})\cdots\delta^{(1)}(\vec{k}_{n})\right\rangle _{{\rm loop}}=-\sum_{j}Z_{ij}\left\langle O_{j}(\vec{k})\delta^{(1)}(\vec{k}_{1})\cdots\delta^{(1)}(\vec{k}_{n})\right\rangle \qquad\forall k_{i}\rightarrow0\label{eq:cond_2}
\end{equation}
where the subscript ``loop'' refers to the diagrams that involve
at least one contraction between the internal legs of the operator
$O_{i}$, whereas the ``tree'' diagrams in Eq.~(\ref{eq:renom_cond})
do not contain such contractions\footnote{For example if there is a composite operator $\left[\delta^{2}\right]_{q}$,
there will be two contractions with ``interaction'' vertices such
that if one of the pair of the interaction vertices are contracted
with an internal line, there will be a loop topology in the convolution.}. Since the operator set has a finite number of elements at any given
perturbative order, the dimensionality of $Z_{ij}(\Lambda)$ is finite
in perturbation theory at a given truncation order. This means that
the number of conditions that needs to be imposed in the form Eq.~(\ref{eq:renom_cond})
to determine $Z_{ij}(\Lambda)$ is finite.

In the case of describing coarse grained fluid density fields, the
$n$th order perturbation theory operators are partly those of SPT which form one susbset of composite operators: e.g.~$\delta(k)=\delta^{(1)}(k)+\delta^{(2)}(k)+...$
with $\delta^{(2)}(k)=\int d^{3}q_{1}d^{3}q_{2}F_{2}(q_{1},q_{2})\delta^{(1)}(q_{1})\delta^{(1)}(q_{2})$.
Other composite operator terms of $\delta(k)$ in \eft include a
derivative expansion of the effective pressure term 
\begin{equation}
\delta^{(c_{1})}(k,z)\approx c^{2}(\Lambda,z)k^{2}\delta^{(1)}(k,z)
\end{equation}
where $c^{2}(\Lambda,z)$ is an effective cutoff-dependent sound speed squared parameter. The function $c^{2}(\Lambda,z)$ is fixed by a couple of conditions that deviate from the condition of Eq.~(\ref{eq:cond_2}).  We write 
\begin{equation}
c^{2}(\Lambda,z)= c_{\Lambda}^{2}(k_{\rm{ren}},\Lambda,z)+c_{\rm{ren}}^{2}(k_{\rm{ren}},z)
\label{eq:subtraction}
\end{equation}
where $k_{\rm ren}$ should not be viewed as a renormalization scale but as a parameter defining the subtraction scheme.\footnote{This usage of the term "renormalization scale" is standard \cite{Hertzberg_2014}.}  The function $ c_{\Lambda}^{2}(k_{\rm{ren}},\Lambda,z)$ is fixed by the choice
\begin{equation}
c_{\Lambda}^{2}(k_{\rm{ren}},\Lambda,z)=\frac{P_{13}(k_{{\rm ren}},\Lambda,z)}{2k_{{\rm ren}}^{2}P_{{\rm 11}}(k_{{\rm ren}},z)}
\end{equation}
with $k_{\rm{ren}}\sim 0.1 h$ Mpc$^{-1}$ in this work.
The 
function $c_{\rm{ren}}^{2}(k_{\rm{ren}},z)$ is obtained by a best-fit to the numerical simulations data.  This in the language of ordinary quantum field theory (QFT) renormalization procedure corresponds indirectly to a judicious choice of the renormalization scale that minimizes the theory error.  Because $k_{\rm ren}$ here corresponds to a definition of a subtraction scheme, the analog of the renormalization scale $\mu$ in QFT does not explicitly appear in Eq.~(\ref{eq:subtraction}).

The bias expansion for
galaxies is also a composite operator expansion. In the notation of
\citep{Desjacques_2018}
\begin{equation}
\delta_{g}(\vec{x})=b_{\epsilon}[\epsilon(\vec{x})]+\sum_{\mathcal{O}}\left\{ b_{\mathcal{O}}+[\epsilon_{\mathcal{O}}(\vec{x})]\right\} [\mathcal{O}(\vec{x})].\label{eq:renormalized_bias_expn}
\end{equation}
where  $\epsilon_{\mathcal{O}}(\vec{x})$ are the stochastic fields
which really represent a class of short distance effects and $b_{\mathcal{O}}$
are coefficients that can absorb divergences based on the chosen renormalization
scheme. The bias parameters also contain cutoff dependences that can
be adjusted according to Eq.~(\ref{eq:renom_cond}) to make the observed
$\delta_{g}$ correlators $\Lambda$-independent.

\section{\label{sec:Galaxy-one-loop-contributions}Galaxy one-loop contributions}

The one-loop contribution to the galaxy two-point correlation function
in $k$-space is given by the following convolutional loop integral
terms taken from \citep{Assassi:2014fva}:

\begin{align}
\mathcal{I}_{\delta^{(2)}\delta^{2}}(k,z) & =2\int_{q}F_{2}\left(\vec{q},\vec{k}-\vec{q}\right)P_{{\rm 11}}(q,z)P_{{\rm 11}}(|\vec{k}-\vec{q}|,z),\\
\mathcal{I}_{\delta^{(2)}\mathcal{G}_{2}}(k,z) & =2\int_{q}\sigma^{2}\left(\vec{q},\vec{k}-\vec{q}\right)F_{2}\left(\vec{q},\vec{k}-\vec{q}\right)P_{{\rm 11}}(q,z)P_{{\rm 11}}(|\vec{k}-\vec{q}|,z),\\
\mathcal{F}_{\mathcal{G}_{2}}(k,z) & =4P_{{\rm 11}}(k,z)\int_{q}\sigma^{2}\left(\vec{q},\vec{k}-\vec{q}\right)F_{2}\left(\vec{k},-\vec{q}\right)P_{{\rm 11}}(q,z),\\
\mathcal{I}_{\delta^{2}\delta^{2}}(k,z) & =2\int_{q}P_{{\rm 11}}(q,z)P_{{\rm 11}}(|\vec{k}-\vec{q}|,z),\\
\mathcal{I}_{\mathcal{G}_{2}\mathcal{G}_{2}}(k,z) & =2\int_{q}\left(\sigma^{2}\left(\vec{q},\vec{k}-\vec{q}\right)\right)^{2}P_{{\rm 11}}(q,z)P_{{\rm 11}}(|\vec{k}-\vec{q}|,z),\\
\mathcal{I}_{\delta^{2}\mathcal{G}_{2}}(k,z) & =2\int_{q}\sigma^{2}\left(\vec{q},\vec{k}-\vec{q}\right)P_{{\rm 11}}(q,z)P_{{\rm 11}}(|\vec{k}-\vec{q}|,z),
\end{align}
where $\int_{q}\equiv\int\frac{d^{3}q}{\left(2\pi\right)^{3}}$, $\sigma^{2}(\vec{k}_{1},\vec{k}_{2})=\left(\vec{k}_{1}\cdot\vec{k}_{2}/k_{1}k_{2}\right)^{2}-1$,
and $F_{2}$ is the mode-coupling kernel for $\delta_{k}^{(2)}$.

\section{\label{sec:Semi-analytic-expression-for-csrenMX}Semi-analytic expression
for $c_{{\rm ren,MX}}^{2}$}

Consider the one-loop matter power spectrum in the EFTofLSS perturbation
theory from Eq.~(\ref{eq:renormalized_1loopPmm}),
\begin{align}
P_{{\rm 1-EFT}}(k,z) & =D^{2}(z)P_{11}(k)+D^{4}(z)\left[P_{{\rm 22}}(k)+P_{{\rm 13}}(k,\Lambda)-2D^{-2}(z)c_{{\rm \Lambda}}^{2}(k_{{\rm ren}},\Lambda,z)k^{2}P_{11}(k)\right]\nonumber \\
 & -2D^{2}(z)c_{{\rm ren}}^{2}(k_{{\rm ren}},z)k^{2}P_{11}(k).\label{eq:1-loopEFTPS}
\end{align}
where $D(z)$ is
the normalized growth function, $D(z)\equiv D_{+}(z)/D_{+}(0)$. We restrict
the above expression to scales where the two-loop and other
higher order contributions are negligible such that $P_{{\rm 1-EFT}}(k)\approx P_{{\rm NL}}(k)$
where $P_{{\rm NL}}(k)$ is the exact non-linear power spectrum. We
consider a renormalization choice where the $\Lambda$-dependent \ctp
$c_{\Lambda}^{2}(k_{{\rm ren}},\Lambda,z)$ in Eq.~(\ref{eq:1-loopEFTPS})
is expressed as:\textbf{
\begin{align}
c_{{\rm \Lambda}}^{2}(k_{{\rm ren}},\Lambda,z) & =\frac{P_{13}(k_{{\rm ren}},\Lambda,z)}{2k_{{\rm ren}}^{2}P_{{\rm 11}}(k_{{\rm ren}},z)}=D^2(z)\frac{P_{13}(k_{{\rm ren}},\Lambda)}{2k_{{\rm ren}}^{2}P_{{\rm 11}}(k_{{\rm ren}})}.\label{eq:c2lambda}
\end{align}
}At the scale $k=k_{{\rm ren}}$, the expression in
Eq.~(\ref{eq:1-loopEFTPS}) reduces to 
\begin{equation}
P_{{\rm 1-EFT}}(k_{{\rm ren}},z)=D^{2}(z)P_{11}(k_{{\rm ren}})+D^{4}(z)P_{{\rm 22}}(k_{{\rm ren}})-2D^{2}(z)c_{{\rm ren}}^{2}(k_{{\rm ren}},z)k_{{\rm ren}}^{2}P_{11}(k_{{\rm ren}}).
\end{equation}
Next we take the difference in the one-loop matter power spectra for
the two different sets of initial conditions: namely, pure adiabatic
(AD), and mixed adiabatic and isocurvature (MX) at $k=k_{{\rm ren}}$
written as
\begin{align}
P_{{\rm 1-EFT}}^{{\rm MX}}(k_{{\rm ren}},z)-P_{{\rm 1-EFT}}^{{\rm AD}}(k_{{\rm ren}},z) & =D^{2}(z)\left(P_{11}^{{\rm MX}}(k_{{\rm ren}})-P_{11}^{{\rm AD}}(k_{{\rm ren}})\right)\nonumber \\
 & +D^{4}(z)\left(P_{22}^{{\rm MX}}(k_{{\rm ren}})-P_{22}^{{\rm AD}}(k_{{\rm ren}})\right)\nonumber \\
 & -2D^{2}(z)k_{{\rm ren}}^{2}\left(c_{{\rm ren,MX}}^{2}(k_{{\rm ren}},z)P_{11}^{{\rm MX}}(k_{{\rm ren}})-c_{{\rm ren,AD}}^{2}(k_{{\rm ren}},z)P_{11}^{{\rm AD}}(k_{{\rm ren}})\right).\label{eq:diff}
\end{align}
In the above expression, the change in the nonlinear matter power
spectrum due to the variation in $c_{{\rm ren}}^{2}$ can be written
as
\begin{equation}
\Delta P_{{\rm NL,c}}(k,z)=-2k_{{\rm ren}}^{2}\Delta c_{{\rm ren}}^{2}(k_{{\rm ren}},z)P_{11}^{{\rm MX}}(k_{{\rm ren}},z)\label{eq:delta_PNL,c}
\end{equation}
where 
\begin{align}
\Delta c_{{\rm ren}}^{2}(k_{{\rm ren}},z) & =c_{{\rm ren,MX}}^{2}(k_{{\rm ren}},z)-c_{{\rm ren,AD}}^{2}(k_{{\rm ren}},z).
\end{align}
Since the parameter $c_{{\rm ren}}^{2}$ is a leading order residual
contribution after integrating out the UV scales, its variation due
to the additional isocurvature power is a nontrivial contribution
from the nonlinear-nonperturbative modes.

By defining the relative fractional difference for a power spectrum
contribution $P_{i}$ as
\begin{equation}
g_{i}(k,z)=\frac{P_{i}^{{\rm MX}}(k,z)-P_{i}^{{\rm AD}}(k,z)}{P_{i}^{{\rm AD}}(k,z)},
\end{equation}
we can write LHS in Eq.~(\ref{eq:delta_PNL,c}) as
\begin{equation}
\Delta P_{{\rm NL,c}}(k,z)=g_{{\rm NL,c}}(k,z)P_{{\rm NL}}^{{\rm AD}}(k,z)\label{eq:delta_PNL,c-2}
\end{equation}
where we take $P_{{\rm NL}}^{{\rm AD}}\approx P_{{\rm 1-EFT}}^{{\rm AD}}$.
Substituting Eq.~(\ref{eq:delta_PNL,c-2}) in Eq.~(\ref{eq:delta_PNL,c}),
we obtain
\begin{equation}
\Delta P_{{\rm {\rm NL,c}}}(k_{{\rm ren}},z)\equiv g_{{\rm NL,c}}(k_{{\rm ren}},z)P_{{\rm NL}}^{{\rm AD}}(k_{{\rm ren}},z)=\left(1+g_{{\rm 11}}(k_{{\rm ren}})\right)\frac{\Delta c_{{\rm ren}}^{2}(k_{{\rm ren}},z)}{c_{{\rm ren,AD}}^{2}(k_{{\rm ren}},z)}P_{c}^{{\rm AD}}(k_{{\rm ren}},z)
\end{equation}
where $g_{11}(k,z)\equiv g_{11}(k)=\Delta P_{11}(k)/P_{11}^{{\rm AD}}(k)$ is obtained
from Eq.~(\ref{eq:Pm_analytical}) as
\begin{equation}
g_{11}(k)=\alpha\left(\frac{f_{c}}{3}\right)^{2}\left(\frac{T_{{\rm iso}}(k)}{T_{{\rm ad}}(k)}\right)^{2}\left(\frac{k}{k_{\mathrm{p}}}\right)^{n_{{\rm iso}}-n_{{\rm ad}}}.
\end{equation}
By factorizing $g_{{\rm NL,c}}(k_{{\rm ren}},z)$ as
\begin{equation}
g_{{\rm NL,c}}(k_{{\rm ren}},z)=-2k_{{\rm ren}}^{2}\,g_{{\rm 11}}(k_{{\rm ren}}){\rm B}(k_{{\rm ren}},n_{{\rm iso}},\alpha,z)
\end{equation}
where the ${\rm B}$ term can either be positive or negative depending
upon the non-linear corrections, we arrive at the expression
\begin{equation}
\Delta c_{{\rm ren}}^{2}(k_{{\rm ren}},z)\approx\frac{P_{{\rm NL}}^{{\rm AD}}(k_{{\rm ren}},z)}{P_{11}^{{\rm AD}}(k_{{\rm ren}},z)}\times\frac{g_{{\rm 11}}(k_{{\rm ren}})}{1+g_{{\rm 11}}(k_{{\rm ren}})}\times{\rm B}(k_{{\rm ren}},n_{{\rm iso}},\alpha,z).\label{eq:dcs_3}
\end{equation}
In the above expression, we note that the first prefactor is isocurvature
independent and is completely determined from the adiabatic component
by replacing $P_{{\rm NL}}^{{\rm AD}}(k_{{\rm ren}},z)$ with $P_{{\rm 1-EFT}}^{{\rm AD}}(k_{{\rm ren}},z)$
and using an empirically measured value of the renormalized fluid
parameter $c_{{\rm ren,AD}}^{2}(k_{{\rm ren}},z)$ from N-body simulations
or data. The second term is analytically computable for a given fiducial
choice of isocurvature model parameters $n_{{\rm iso}}$ and $\alpha$
and measures the fraction of linear power carried by the isocurvature
modes. The ${\rm B}$ term captures the difference in the power for
the density fluctuations between the mixed isocurvature (MX) and pure
adiabatic (AD) from backreaction of nonperturbative UV modes on large
scales and can be obtained by fitting to the numerical results from
N-body simulations. Hence, $B\neq0$ is an indirect measure of the
relative difference in the nonlinear corrections from the small scales
to the coarse-grained effective fluid due to the additional contribution
from primordial blue isocurvature fluctuations.

In order to complete our estimation of $c_{{\rm ren,MX}}^{2}(k_{{\rm ren}},z)$,
we give the following empirical fitting function for ${\rm B}$ at
$k_{{\rm ren}}=0.1{\rm h/Mpc}$ by matching with the data from N-body
simulations:
\begin{align}
{\rm B}(k_{{\rm ren}}=0.1{\rm h/Mpc},n_{{\rm iso}},\alpha,z) & =0.272\times D(z)^{0.858}\alpha^{-0.147}\left(n_{{\rm iso}}-n_{{\rm ad}}\right)\left(\frac{k_{s}}{k_{{\rm p}}}\right)^{n_{{\rm iso}}-n_{{\rm ad}}}
\end{align}
where
\begin{equation}
k_{s}\approx0.2174\,{\rm Mpc^{-1}}
\end{equation}
represents a fixed scale.

Substituting the above empirical expression into Eq.~(\ref{eq:dcs_3})
we obtain the result 
\begin{align}
\Delta c_{{\rm ren}}^{2}(k_{{\rm ren}}=0.1{\rm h/Mpc},z) & \approx\frac{P_{{\rm 1-EFT}}^{{\rm AD}}(k_{{\rm ren}},z)}{P_{11}^{{\rm AD}}(k_{{\rm ren}},z)}\times\frac{g_{{\rm 11}}(k_{{\rm ren}})}{1+g_{{\rm 11}}(k_{{\rm ren}})}\times\nonumber \\
 & \left(0.272\,D(z)^{0.858}\alpha^{-0.147}\left(n_{{\rm iso}}-n_{{\rm ad}}\right)\left(\frac{k_{s}}{k_{{\rm p}}}\right)^{n_{{\rm iso}}-n_{{\rm ad}}}\right)\label{eq:dcs2_final}
\end{align}
for the estimation of renormalized fluid parameter $c_{{\rm ren,MX}}^{2}=c_{{\rm ren,AD}}^{2}+\Delta c_{{\rm ren}}^{2}$
for mixed isocurvature initial conditions for fiducial values of $O(0.01)\lesssim\alpha<1$
and $2\leq n_{{\rm iso}}\leq4$ for redshift range $z\leq2$. The
value of $\Delta c_{{\rm ren}}^{2}$ at a different renormalization
scale can be obtained from the following expression:
\begin{align}
\Delta c_{{\rm ren}}^{2}(k_{{\rm ren,1}},z) & =\Delta c_{{\rm ren}}^{2}(k_{{\rm ren}},z)+\left(\Delta c_{{\rm \Lambda}}^{2}(k_{{\rm ren}},\Lambda,z)-\Delta c_{{\rm \Lambda}}^{2}(k_{{\rm ren,1}},\Lambda,z)\right)
\end{align}
where $k_{{\rm ren}}$ and $k_{{\rm ren,1}}$ are the old and new
renormalization scales respectively, and $\Delta c_{{\rm \Lambda}}^{2}(k_{{\rm ren}},\Lambda,z)\equiv c_{{\rm \Lambda,MX}}^{2}(k_{{\rm ren}},\Lambda,z)-c_{{\rm \Lambda,AD}}^{2}(k_{{\rm ren}},\Lambda,z)$
is the difference in the square of the $\Lambda$-dependent \ctp as defined in Eq.~(\ref{eq:c2lambda}).

\section{\label{sec:Fisher-matrix-and}Fisher matrix and theoretical error
covariance}

For the purpose of Fisher forecasting, we consider a Gaussian likelihood
function \citep{dodelson2020modern}:
\begin{equation}
\mathcal{L}=\frac{1}{\sqrt{\left(2\pi\right)^{N_{c}}|C|}}\exp\left[-\frac{1}{2}\left(d-t\right)^{{\rm T}}C^{-1}\left(d-t\right)\right]\label{eq:likelihood_defn}
\end{equation}
where $N_{c}$ is the total number of momentum configurations (for
the power spectrum (bispectrum), $N_{c}$, is the total number of
momentum bins (triangle configurations)). Here, $d$ and $t$ are
the data and theoretical prediction vectors while $C$ is the covariance
matrix which is to be taken as the sum of data covariance and the
theoretical covariance \citep{Baldauf:2016sjb,dodelson2020modern}.
Using the likelihood function, the Fisher information matrix is defined
as the curvature of the likelihood at the fiducial parameter set along
each model parameter $p_{n}$:
\begin{equation}
F_{ij}=-\left.\left\langle \frac{\partial^{2}\ln\mathcal{L}}{\partial p_{i}\partial p_{j}}\right\rangle \right|_{p=p_{{\rm fid}}},\label{eq:fisher_defn}
\end{equation}
where the matrix is evaluated at a fiducial point in parameter space
$p_{{\rm fid}}$ and $\left\langle ...\right\rangle $ implies taking
an ensemble average. The Fisher matrix then allows one to understand
how well various parameters of the theoretical model can be constrained.
The unmarginalized error on one single parameter $p_{i}$ is estimated
by the inverse of the respective diagonal element in the Fisher matrix
\begin{equation}
\sigma(p_{i})=\frac{1}{\sqrt{F_{ii}}}.\label{eq:umg_sigma}
\end{equation}
On the other hand, one can obtain marginalized error estimate by integrating
the probability over all other model parameters. This is easily obtained
from the Fisher information matrix as
\begin{equation}
\sigma(p_{i})=\sqrt{\left(F^{-1}\right)_{ii}}.\label{eq:mg_sigma}
\end{equation}
For marginalization over a smaller set of parameters, we refer the
reader to \citep{Coe:2009xf}. From the definitions given in Eqs.~(\ref{eq:likelihood_defn})
and (\ref{eq:fisher_defn}) it follows that the power spectrum Fisher
matrix is \citep{dodelson2020modern}:
\begin{equation}
F_{ij}^{s}=\sum_{z_{l}}\sum_{k,k'}\frac{\partial P_{s}\left(k,z_{l}\right)}{\partial p_{i}}\left(C^{-1}(z_{l})\right)_{kk'}^{s}\frac{\partial P_{s}\left(k',z_{l}\right)}{\partial p_{j}}
\end{equation}
where the sum runs over all measured binned redshifts and momentum
bins and $P_{s}$ is the theoretical power spectrum where $s$ characterizes
the spectrum type which we can be either matter or biased tracer like
galaxy, halo or mixed matter-galaxy. All terms are evaluated at the
fiducial values of the model parameters. Here we will consider that
the redshift bins are wide enough such that the cross spectra between
the bins vanish (see discussion in \citep{LESGOURGUES_2006} and Chapter
14 of \citep{dodelson2020modern} for details). Also see \citep{Bellomo:2020pnw}
for the situations where these assumptions might fail.

The power spectrum covariance matrix $C_{kk'}^{s}$ can be written
as sum of data covariance ($C_{d,kk'}^{s}(z_{l})$) and theory error estimate covariance ($C_{e,kk'}^{s}(z_{l})$) \citep{Baldauf:2016sjb}:
\begin{equation}
C_{kk'}^{s}(z_{l})=C_{d,kk'}^{s}(z_{l})+C_{e,kk'}^{s}(z_{l}).
\end{equation}
The data covariance for the galaxy power spectrum is
\begin{equation}
C_{d,kk'}^{s}(z_{l})=\frac{\left(2\pi\right)^{3}}{V(z_{l})}\frac{f_{{\rm sky}}^{-1}}{2\pi k^{2}k_{{\rm bin}}}\left(P_{s}(k,z_{l})+P_{s,{\rm shot}}(z_{l})\right)^{2}\delta_{kk'}\label{eq:CD}
\end{equation} 
where $V(z_{l})$ is the volume of the shell centered at $z_{l}$
with the bin width $\Delta z_{l}$ and is given by the expression
\begin{equation}
V(z_{l})=\frac{4\pi}{3}\left[r^{3}\left(z_{l}+\frac{\Delta z}{2}\right)-r^{3}\left(z_{l}-\frac{\Delta z}{2}\right)\right],
\end{equation}
$r(z')$ is the comoving distance to the redshift $z'$, $f_{{\rm sky}}$
is the observed fraction of sky, $k_{{\rm bin}}$ is the momentum
bin-width and $P_{{\rm shot}}$ is the contribution from the shot-noise
which is equal to $1/\bar{n}(z_{l})$ in case of galaxies where $\bar{n}$
is the average number density of galaxies\footnote{By construction, the Fisher matrix is expected to give inaccurate
results for non-Gaussian posteriors. It is well known that the use
of a Gaussian approximation for the data covariance matrix $C_{kk'}$,
at scales comparable to or much smaller than nonlinear scale, may break
down due to loop corrections, one-halo, and perturbation non-diagonal
terms. A more careful analysis, in this case, requires a full MCMC forecasting,
although a few improvements have been suggested by taking into account
non-Gaussian posteriors \citep{Joachimi:2011iq,Sellentin:2014zta,Sellentin:2015axa,Amendola:2016wim,Wolz_2012}.
Interestingly, the analysis of \citep{Wadekar:2020hax} suggests that
inclusion of non-Gaussian covariance (with regular trispectrum and
super-sample covariance effects) has marginal $<10\%$ effect on parameter
error bars.}. The theoretical error covariance matrix as defined in \citep{Baldauf:2016sjb}
is given as
\begin{equation}
C_{e,kk'}^{s}(z_{l})=E_{k}^{s}\exp\left[-\frac{\left(k-k'\right)^{2}}{2\Delta k^{2}}\right]E_{k'}^{s}\label{eq:CE}
\end{equation}
where $E_{k}^{s}\equiv E^{s}(k,z)$ is the theoretical error estimate
(smoothed envelope without any fast variation as a function of wave
vectors) for the spectrum type $s$. The explicit expressions for
$E^{s}(k,z)$ are given in Sec.~\ref{subsec:Fiducial-parameter-values}.
Note that in the above expression for $C_{e}^{s}$, the errors in
various momentum bins are correlated through a Gaussian correlation
with a common width set by $\Delta k$. In our analysis we found that
a choice of $0.05<\Delta k\left({\rm h/Mpc}\right)^{-1}<0.13$ does
not lead to significant changes $(\lesssim10\%)$ in the error estimates
from Fisher analysis. For a more detailed discussion refer to \citep{Baldauf:2016sjb,Chudaykin:2019ock}.
Similarly, for bispectrum, the Fisher information matrix is given
as

\begin{equation}
F_{ij}^{b}=\sum_{z_{l}}\sum_{T,T'}\frac{\partial B_{s}\left(T,z_{l}\right)}{\partial p_{i}}\left(C^{-1}(z_{l})\right)_{TT'}^{s}\frac{\partial B_{s}\left(T',z_{l}\right)}{\partial p_{j}}
\end{equation}
where the sum runs over all possible triangle configurations such
that $\vec{k}_{1}+\vec{k}_{2}+\vec{k}_{3}=0$ with the following ordering
of the scales $k_{1}\geq k_{2}\geq k_{3}$. Thus, 
\begin{equation}
\sum_{T}\equiv\sum_{k_{1}=k_{\min}}^{k_{\max}}\sum_{k_{2}=k_{\min}}^{k_{1}}\sum_{k_{3}=k_{*}}^{k_{2}}
\end{equation}
where $k_{*}=\max\left(k_{\min},k_{1}-k_{2}\right)$. The total bispectrum
covariance matrix is given as
\begin{equation}
C_{TT'}^{s}(z_{l})=\delta_{TT'}\frac{\left(2\pi\right)^{3}}{V(z_{l})}s_{123}\frac{\pi f_{{\rm sky}}^{-1}}{k_{{\rm bin}}^{3}}\frac{1}{k_{1}k_{2}k_{3}}\prod_{j=1}^{3}\left(P_{s}(k_{j},z_{l})+P_{s,{\rm shot}}(z_{l})\right)+\left(C_{e}^{s}(z_{l})\right)_{TT'}
\end{equation}
where $s_{123}$ is the symmetry factor equal to $1,2$ or $6$ for
general, isosceles, and equilateral triangles respectively.

While the data covariance falls as $1/k^{2}$ for small scales, the
theory error covariance tends to increase rapidly as we approach scale,
$k_{{\rm Th}}(z)$, where the underlying theory tends to break down.
Hence, the signal to noise ratio is maximum close to the minimum of
the sum of the two covariances. Therefore, by including theoretical
error within the covariance matrix the signal is naturally cutoff
at scales $\approx k_{{\rm Th}}(z)$.

Combining the power spectrum and bispectrum information is equivalent
to summing up the Fisher matrices
\begin{equation}
F=F_{p}+F_{b}+{\rm diag\left(1/\sigma_{p_{i}}^{2}\right)}
\end{equation}
where $\sigma_{p_{i}}$ is the assumed prior on a parameter $p_{i}$.

\section{Comparison to Halofit}
\label{sec:halofit}

\begin{figure}
\begin{centering}
\includegraphics[scale=0.5]{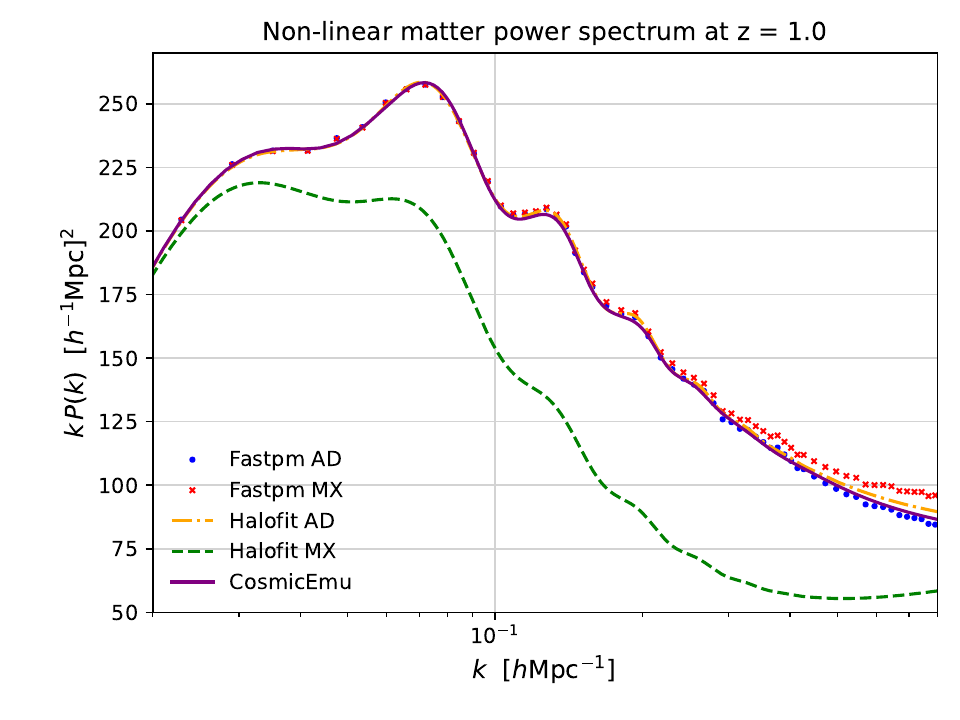}~\includegraphics[scale=0.5]{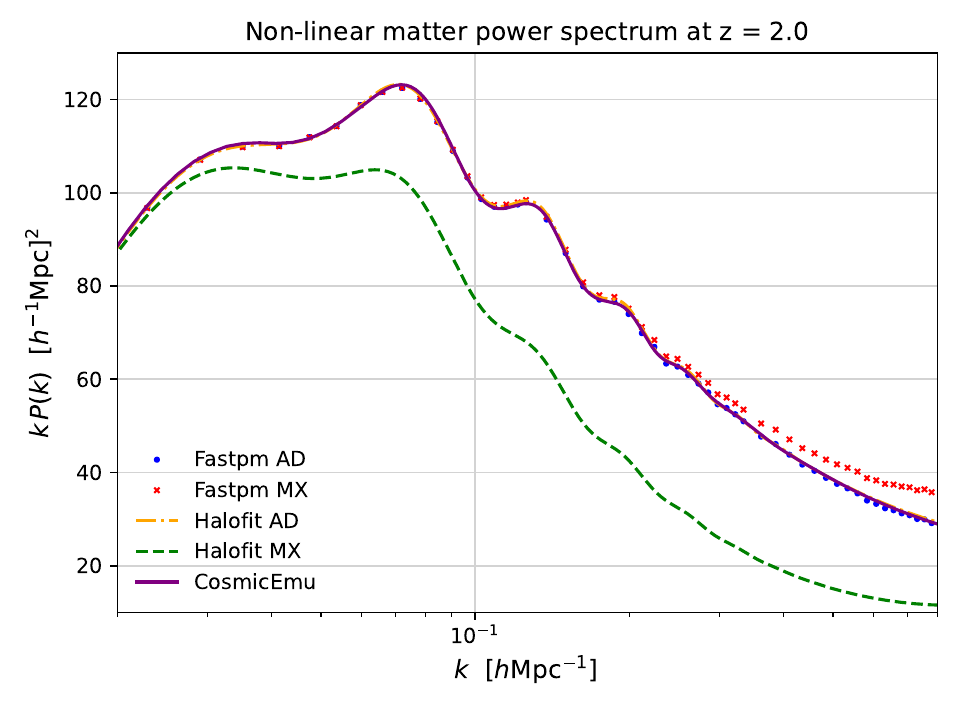}
\par\end{centering}
\caption{\label{fig:Halofit}Performance of Halofit (from CLASS
code \citep{Blas:2011rf}) compared to N-body data (from FastPM \citep{Feng:2016yqz})
for predicting the matter power spectrum for pure adiabatic (AD) and mixed
(MX) ICs. For comparison we also plot the curve from CosmicEmu \citep{Moran:2022iwe} which is only available for pure adiabatic case. We highlight that Halofit fits N-body data well for
adiabatic ICs while grossly misrepresenting the power on mildly nonlinear and even large
scales for mixed state. To compare with the performance of EFT we refer the readers to Fig.~\ref{fig:eft-fastpm-compare}. This figure has been constructed at redshift
$z=1$ and $z=2$ with an uncorrelated CDM isocurvature fraction $\alpha=0.25$
for a high blue spectral index $n_{{\rm iso}}=3.75$. 
}
\end{figure}

Some forecasts use theoretical predictions for non-linear galaxy clustering generated with Halofit \citep{Takahashi:2012em}. However, it was shown in \citep{Reimberg:2018kwn} that because the Halofit is not sufficiently accurate in reproducing the derivatives of the power spectrum, other methods should be used for Fisher forecasts. Furthermore, as can be seen in Fig.~\ref{fig:Halofit}, if we apply Halofit naively without adjusting its coefficients to isocurvature (by fitting them to simulations with isocurvature), it grossly under-predicts the power on BAO scales for a mixed state containing a sizable fraction of blue isocurvature power, as we now explain. Halofit relies on the Halo model \cite{Takahashi:2012em} for its fitting form, and the mismatch between the fit and the data appears in the two-halo term which governs the large-scale behavior. This may be explained if we note that the two-halo term in Halofit is evaluated using a set of fitting parameters, ($\alpha_n,\,\beta_n$), that are given by polynomials as a function of the effective spectral index $n_{\rm{eff}}$ and the curvature $C$ of the linear variance $\sigma^2(R)$ evaluated at the scale $k_\sigma$ where the $\sigma^2(1/k_\sigma) = 1$: the two-halo term contribution to the power spectrum is
\begin{equation}
\Delta_Q^2(k)=\Delta_L^2(k)  \frac{\left(1+\Delta_L^2(k) \right)^{\beta_n}}{1+\alpha_n\Delta_L^2(k)} e^{-f(k/k_{\sigma})}    
\end{equation} where $\Delta_L^2$ is the dimensionless linear power spectrum, $f(y)=y/4+y^2/8$, and
\begin{eqnarray}
\alpha_n & = & \left| 6.0835 + 1.3373 n_{\mathrm{eff}}-0.1959 n_{\mathrm{eff}}^2- 5.5274 C \right| \label{eq:alphan_halo} \\
\beta_n & = & 2.0379-0.7354 n_{\mathrm{eff}} + 0.3157 n_{\mathrm{eff}}^2 + 1.2490 n_{\mathrm{eff}}^3 + 0.3980 n_{\mathrm{eff}}^4-0.1682C
\end{eqnarray}
The coefficients such as 5.5274 of Eq.~\ref{eq:alphan_halo} of the polynomial functions were obtained by fitting the halo model to a large set of $\Lambda$CDM simulations without any blue isocurvature amplitude. Hence, it may be argued that the fitting coefficients are not well suited to give an accurate prediction for the non-linear power spectrum with a sizable isocurvature fraction. To illustrate the difference, we found that the value of the parameter set $\left( n_{\rm{eff}},\,C \right) $ for the pure adiabatic and our fiducial mixed scenario (with $n_{\rm{iso}} = 3.75$ and $\alpha = 0.25$) at redshift $z=1$ are $(-1.993,\, 0.252)$ and $(-1.547,\, -0.333)$ respectively. Note that apart from a relatively large difference in the effective spectral index, the curvature of the linear power spectrum has opposite signs for the two cases. Since the curvature for the mixed case is negative, it results in a relatively larger magnitude of $\alpha_n$ compared to the adiabatic case which leads to a larger suppression of the two-halo term.  This explains qualitatively the suppression of the naive application of the Halofit.  One way to correct this would be to obtain a new set of numerical coefficients appearing in $\alpha_n$ based on matching to mixed initial condition simulations.

\bibliographystyle{JHEP2}
\bibliography{ref2,blu_iso_axion_cdm,ref,eftlss}

\end{document}